\documentclass[useAMS,amssymb,epsfig,usegraphicx,usenatbib,twocolumn,10pt]{aastex63}

\newcommand{\angstrom}{\mbox{\normalfont\AA}}

\usepackage{amsmath}	% Advanced maths commands
\usepackage{amssymb}	% Extra maths symbols

\received{-}
\revised{-}
\accepted{-}
\submitjournal{ApJ}

\shorttitle{The UDISG in the Bootes Void}
\shortauthors{Pandey et al.}
\graphicspath{{./}{figures/}}
\begin{document}

\title{The Ultraviolet Deep Imaging Survey of Galaxies in the Bootes Void I: catalog, color-magnitude relations and star formation}%\footnote{Released on \today}}

\correspondingauthor{Divya Pandey}
\email{divya\_pandey@nitrkl.ac.in}
%in square bracket write orchid id
\author{Divya Pandey}
\affiliation{Department of Physics and Astronomy,
National Institute of Technology,
Rourkela 769 008, India}

%\collaboration{1}{(AAS Journals Data Scientists collaboration)}

%\author{Butler Burton}
%\affiliation{Leiden University}
%\affiliation{AAS Journals Associate Editor-in-Chief}
\author{Kanak Saha}
\affiliation{Inter-University Centre for Astronomy \& Astrophysics,
Postbag 4, Ganeshkhind, Pune 411 007, India
}

%\nocollaboration{1}

\author{Ananta C. Pradhan}
\affiliation{Department of Physics and Astronomy,
National Institute of Technology,
Rourkela 769 008, India}




\begin{abstract}

We present a deep far and near-ultraviolet (FUV and NUV) wide-field imaging survey of galaxies in the Bootes Void using Ultra-Violet Imaging Telescope onboard {\em AstroSat}. Our data reach $5\sigma$ limiting magnitudes for point sources at 23.0 and 24.0 AB mag in FUV and NUV respectively. We report a total of six star-forming galaxies residing in the Bootes Void alongside the full catalog, and of these, three are newly detected in our FUV observation.

Our void galaxy sample spans a range of UV colors $(-0.35\, \leq$ FUV$-$NUV $\leq\, 0.68)$ and absolute magnitudes $(-14.16\,  \leq\, \mathrm{M_{NUV}}\, \leq\, -18.65)$. In addition, {\em Sloan Digital Sky Survey} and {\em Two-micron All Sky Survey} archival data are being used to study UV, optical, and infrared color-magnitude relations for our galaxies in the void. We investigate the nature of bi-modal color distribution, morphologies, and star formation of the void galaxies. Most of the galaxies in our sample are fainter and less massive than L$^{\ast}$ galaxies, with M$_\mathrm{r} > -20$. Our analysis reveals a dominant fraction of bluer galaxies over the red ones in the void region probed. The internal and Galactic extinction corrected FUV star formation rates (SFRs) in our void galaxy catalog varies in a large range of $0.05$ to $51.01$ M$_{\odot} yr^{-1}$, with a median $3.96$ M$_{\odot} yr^{-1}$. We find a weak effect of the environment on the SFRs of galaxies. Implications of our findings are discussed.  

\end{abstract}



\section{INTRODUCTION}
Cosmic web, largely composed of voids, filaments and wall-like structures, is observed to be inhomogeneous at mega-parsec scale. The large-scale structures that we see in our present day universe are a manifestation of the primordial gravitational density fluctuations \citep{1993PhRvD..47.1333S}. The cosmic voids occupy $\sim$77\% of the cosmic volume and they represent $\sim$15\% of the total halo mass content implying that the average density of void is around 20\% of the average cosmic density \citep{Cautun_2014}. The formation and evolution of the voids depend on two processes, i$.$e$.$, small voids merge to shape into a larger under density and due to collapse of over-densities around a region in space \citep{2004MNRAS.350..517S,2016IAUS..308..493V}. They usually tend to exist within the cosmic web \citep{Libeskindetal2018} in a spherical foam-like structure. Such voids can be populated by substructures such as mini-sheets and filaments that run through the voids. As these voids grow older they become progressively empty and possess less substructures within them \citep{Sahnietal1994}.

Typical size of a large scale void ranges from 20h$^{-1}$ to 50h$^{-1}$ Mpc, but its depth remains unclear with an under-density of ${\delta}_\mathrm{{v}} = \frac{{\rho}_v}{<{\rho}>}-1 \approx -0.8$. The voids were first discovered in observation by \citet{1978ApJ...222..784G,1978MNRAS.185..357J}. Later \citet{1981ApJ...248L..57K,1987ApJ...314..493K} discovered the Bootes Void, one of the largest void present in the northern hemisphere. Much recently, {\em Sloan Digital Sky Survey (SDSS)} provided a detailed structure of cosmic voids using large-scale structure galaxy catalog from {\em Baryon Oscillation Spectroscopic Survey (BOSS)} \citep{Mao_2017}.  
\par 
Voids are thought to provide a pristine environment for understanding the secular evolution and dynamics of galaxies \citep{VandeWeygaert_Platen2011} as they are devoid of phenomena typically active in a denser medium like groups and clusters, e$.$g$.$, ram pressure stripping \citep{ram_pressure}, gas strangulation \citep{2015Natur.521..192P} (\textit{galaxy nurture}). As a result, the evolution of void galaxies is thought to be slower than those in denser medium leading to an abundance of young galaxies at primitive stages of evolution, thus, studying void galaxies may unearth key features of the early stages of galaxy formation scenario. In fact, a number of objective prism surveys \citep{1987ApJ...314L..33M}, imaging and spectroscopic observations \citep{2002AJ....123..142C,1996AJ....111.2150S,1995AJ....109..981W} of the Bootes Void show remarkable similarity in the overall void galaxy properties. The galaxies present in voids tend to be bluer than wall and field galaxies with large specific star formation rates (sSFRs) \citep{Moorman_2016}. Voids are mainly populated with late-type galaxies although the presence of active galactic nuclei and early-type galaxies have also been reported recently \citep{2017MNRAS.464..666B}. Not only that, void galaxies have also shown evidences of recent merger interaction \citep{2016IAUS..308..591K,Grogin_2000}. Some of these galaxies show unusual morphological features such as knots, asymmetries, apparent one-spiral arm, and offset nucleus \citep{1997}. Based on these recent reports, one would expect to find a complete spectrum of galaxy morphology at different evolutionary phases in a void environment.

It has been shown that the global properties such as morphology, color, and star formation rates (SFRs) of galaxies depend predominantly on their local environment and internal driven mechanism rather than their global environment \citep{1996AJ....111.2150S,Tinker_2009,2015MNRAS.453.3519P}. Based on a comparative analysis of the emission line galaxies (ELGs) situated in sparse and dense environments, it has been suggested that the local environment density around a galaxy may have no effect on its chemical evolution \citep{2019ApJ...883...29W}. On the other hand, a void environment has been shown to affect the size and stellar masses of the galaxies inside them \citep{2017MNRAS.464..666B}. The evolutionary history of galaxies is shown to depend on whether a galaxy resides deep inside the void or on the periphery as well as on the size of the host void \citep{Ricciardelli_2014,Ricciardelli_2017}. It remains unclear at what scale and which galaxy properties are affected by the environment which would require multi-wavelength deep imaging and spectroscopic surveys of the void region. {\em SDSS} has already done a great job in this aspect. Deep imaging observation of the void in the far and near ultraviolet (FUV and NUV) bands with {\em SDSS} like spatial resolution is, however, missing since {\em Galaxy Evolution Explorer (GALEX)} \citep{2005ApJ...619L...1M} did shallow surveys ($\sim 205$ sec) of the void region (e$.$g$.$, of the Bootes Void in the northern hemisphere).

In this work, we present a deep imaging survey of about 615~sq. arcmin region of the Bootes Void in FUV and NUV filters of the Ultra-Violet Imaging Telescope (UVIT) on-board  {\em AstroSat} Satellite \citep{2017AJ....154..128T}. Since the galaxies in voids show bluer color and are star-forming, the stellar population would contain a significant fraction of young stars (O, B-type) of intermediate masses (2-5 M$_{\odot}$ which actively emit in FUV and NUV \citep{Grogin_1999,rojasss,rojasPS}. The FUV emission in a galaxy arises from the photosphere of massive O- and B-type stars and therefore, it traces star formation going on in a galaxy over a timescale of ${\sim}$ 10$^{8}$ yr \citep{calzetti2012star,Lee_2009}. Using our UV deep imaging survey, we produce a catalog of void galaxies with fluxes from FUV to near-infrared (NIR) and study their morphology, UV-optical and NIR color-magnitude relations \citep{Wyder_2007,2007ApJS..173..185G,2004ApJ...600..681B}, their star formation properties and compare them with non-void galaxies.

\par
The work is organized as follows: in section~\ref{sec:observation}, we describe our data used for analysis, and in section~\ref{sec:DRA}, we explain the procedure adopted for data reduction and analysis. We briefly discuss our methodology for catalog preparation and photometric redshift calculation in section~\ref{sec:CP} whereas in section~\ref{sec:detections}, we deduce the reliability of our UVIT detection followed by section~\ref{sec:DE}, \ref{sec:Mstar} and \ref{sec:sfr}, where we discuss about the internal dust obscuration, stellar masses and quantify UV emission of galaxies. In section~\ref{sec:cmd}, we discuss properties of void galaxies detected in the analysis on the basis of various color-magnitude diagrams (CMDs). Finally, in section~\ref{sec:discuss}, we conclude our findings and examine the future prospects of our research. 
\par
A standard $\Lambda$CDM cosmology with ${\Omega}_M$ = 0.3, ${\Omega}_{\Lambda}$ = 0.7, and H$_{0}$ = 70 km s$^{-1}$ Mpc$^{-1}$ is assumed in this work. 
AB magnitude system \citep{1974ApJS...27...21O} has been followed throughout the work. We have converted {\em 2MASS} Vega magnitudes to AB using conversions given in \citet{Blanton_2005} to maintain uniformity of the magnitude scale. 

\begin{figure}

	\includegraphics[width=\columnwidth]{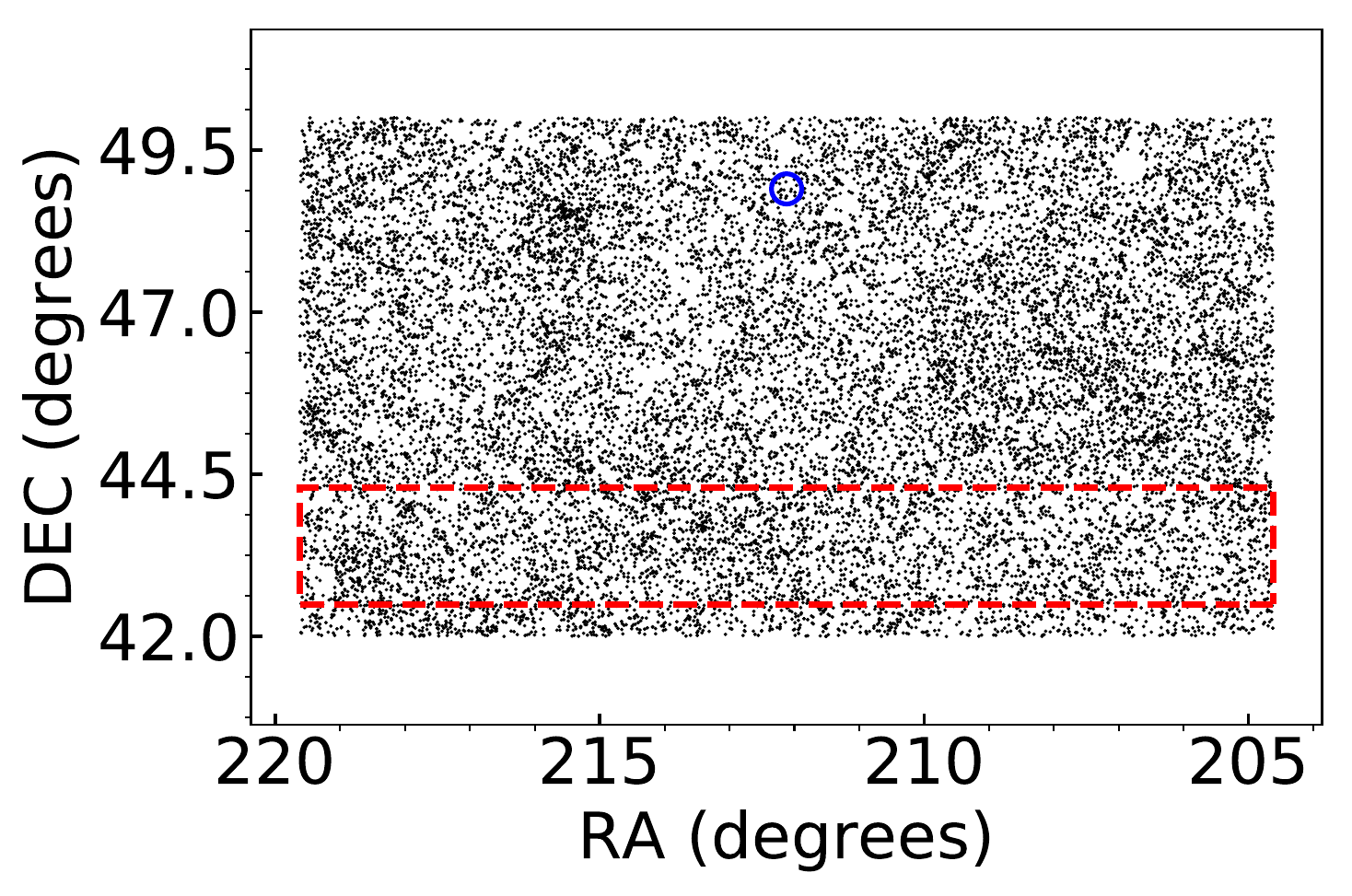}
    \caption{Sky map comprises of a section in the Bootes Void. The UVIT observation covers the area marked by the blue circle. The red open rectangle denotes the area observed by the {\em KISS} survey \citep{2019ApJ...883...29W}. The rectangular area extends to RA range 178.5$^{\circ}$ – 244.4$^{\circ}$, DEC range 42.55$^{\circ}$ – 44.35$^{\circ}$. Each black dot represents a galaxy from the {\em SDSS} archival catalog.}
    \label{fig:KISS survey}
\end{figure}

\begin{figure*}
    \centering
    \includegraphics[width=0.45\linewidth]{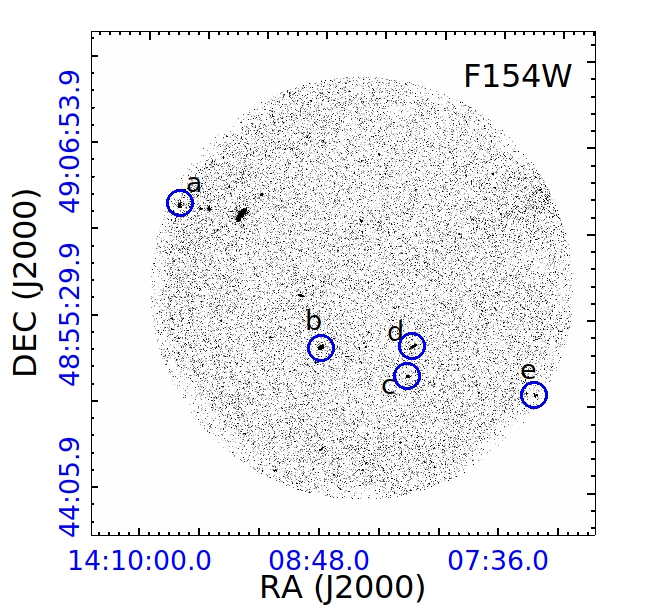}
    \includegraphics[width=0.45\linewidth]{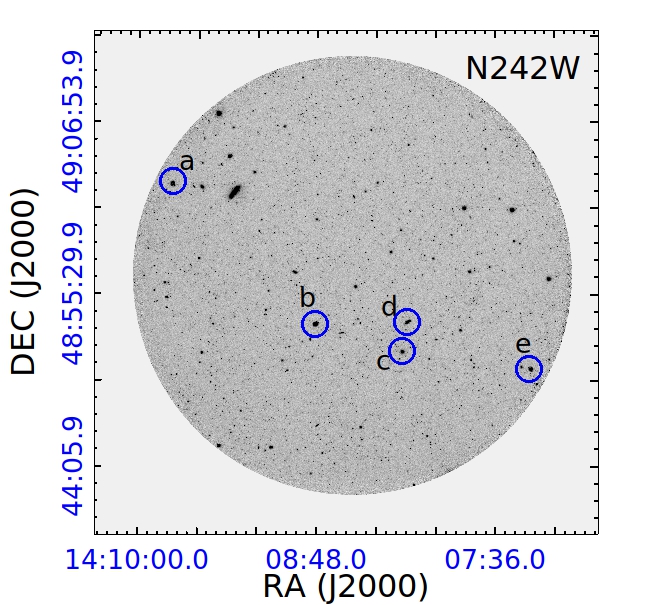}\\
    \includegraphics[width=0.9\linewidth]{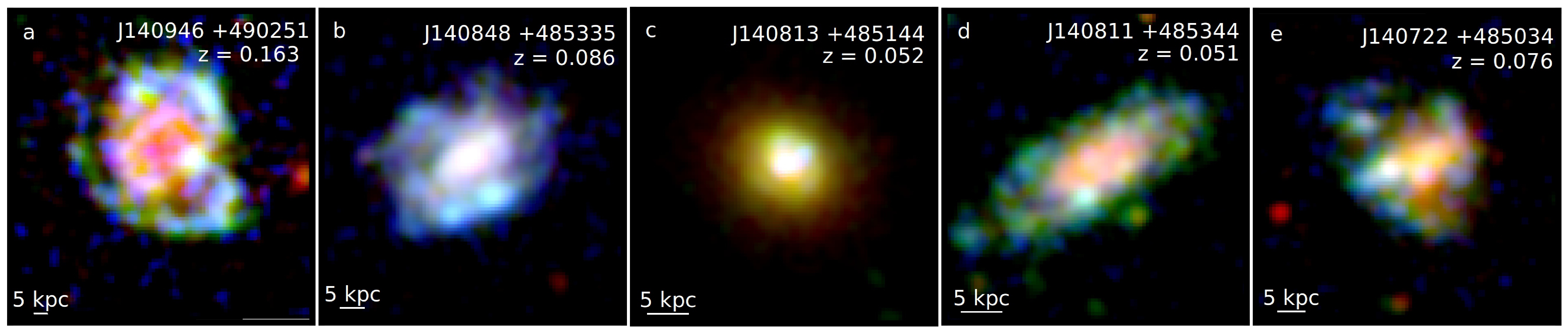}

    \caption{Top panel: Bootes Void observed in the FUV and NUV filters of {\em AstroSat}/UVIT centered at $(\alpha, \delta)\, =\, (14^h\ 08^m\ 27.8^s,\ +48^d\ 55^m\ 56.6^s)$ with a diameter of 28$^{\prime}$ each. Bottom panel: Color composite images of some peculiar galaxies in the Bootes Void region marked with alphabets in the FoV; (red: {\em SDSS} r-filter, green: UVIT NUV filter, blue: UVIT FUV filter).}
    \label{fig:FOV}
\end{figure*}

\section{{\em AstroSat} OBSERVATION and other archival data }
\label{sec:observation}
The {\em AstroSat} is India's first dedicated multi-wavelength space satellite launched by Indian Space Research Organization (ISRO) in September, 2015. UVIT on-board {\em AstroSat} satellite observes primarily in three channels: FUV (${\lambda}=$ 1300 - 1800 ${\angstrom}$), NUV (${\lambda=}$ 2000 - 3000 ${\angstrom}$) and visible (${\lambda}=$ 3200 - 5500 ${\angstrom}$) wavelength bands. The field of view (FoV) of each channel is about 28$^\prime$ in diameter with pixel size of $\approx 0^{\prime\prime}.417$ and spatial resolution of $< 1^{\prime\prime}.8$ in FUV and NUV channels  \citep{2017AJ....154..128T}. The angular resolution of UVIT is 3-4 times higher as compared to previously launched UV space telescope {\em GALEX}.
\par We proposed to explore an area of 615~sq$.$ arcminutes in the Bootes Void to observe with UVIT. The aforementioned area is centered at $\alpha$ = 212.115$^{\circ}$/ 14$^h$08$^m$27.8$^s$ and $\delta$ = 48.932$^{\circ}$/ 48$^d$55$^m$56.6$^s$.
Observations were taken in BaF2 ($\lambda_\mathrm{eff} = 1541\, {\angstrom}$) and Silica-1 ($\lambda_\mathrm{eff} = 2418\, {\angstrom}$) filters of UVIT. The total on-source exposure time assigned to BaF2 (F154W) and Silica-1 (N242W) filters $\approx$ 10000~sec each (PI: Kanak Saha). Figure~\ref{fig:KISS survey} shows the field of observation of a recent {\em KPNO International Spectroscopic Survey (KISS)} of ELGs in direction of Bootes Void \citep{2019ApJ...883...29W} and our UVIT FoV. The top panel of Figure~\ref{fig:FOV} shows FUV/NUV FoV observed by UVIT; in the bottom panel, we have shown the color composite images of five peculiar galaxies detected in our FoV. The images are color coded as follows- red: {\em SDSS} r filter, green : UVIT NUV , and blue: UVIT FUV. Two of the five galaxies (third and fourth image) in the figure are void members while the others lie outside the void. \par
To extend this survey further into the optical and infrared parts of the electromagnetic spectrum, we have included {\em SDSS} and {\em 2MASS} observations. {\em SDSS} has five filters \textit{u}, \textit{g}, \textit{r}, \textit{i}, \textit{z} having mean wavelengths  3560 ${\angstrom}$, 4680 ${\angstrom}$, 6180 ${\angstrom}$, 7500 ${\angstrom}$, 8870 ${\angstrom}$, respectively, spanning over the optical and infrared bands of the electromagnetic spectrum \citep{2000AJ....120.1579Y}. We have used well calibrated archival imaging data from the {\em SDSS} Data Release 12 (DR12) \citep{2015ApJS..219...12A} for the same patch of sky in the Bootes Void as observed by UVIT. Similarly, imaging archival data from {\em 2MASS} have also been taken up which observes in NIR wavelength filters \textit{J} ,\textit{H}, \textit{Ks} with mean wavelengths 1.24 ${\mu}$m, 1.66 ${\mu}$m, 2.16 ${\mu}$m, respectively \citep{2006AJ....131.1163S}. 

%%%%%%%%%%%%%%%%%%

\begin{figure*}
    \centering
    \includegraphics[width=0.43\linewidth]{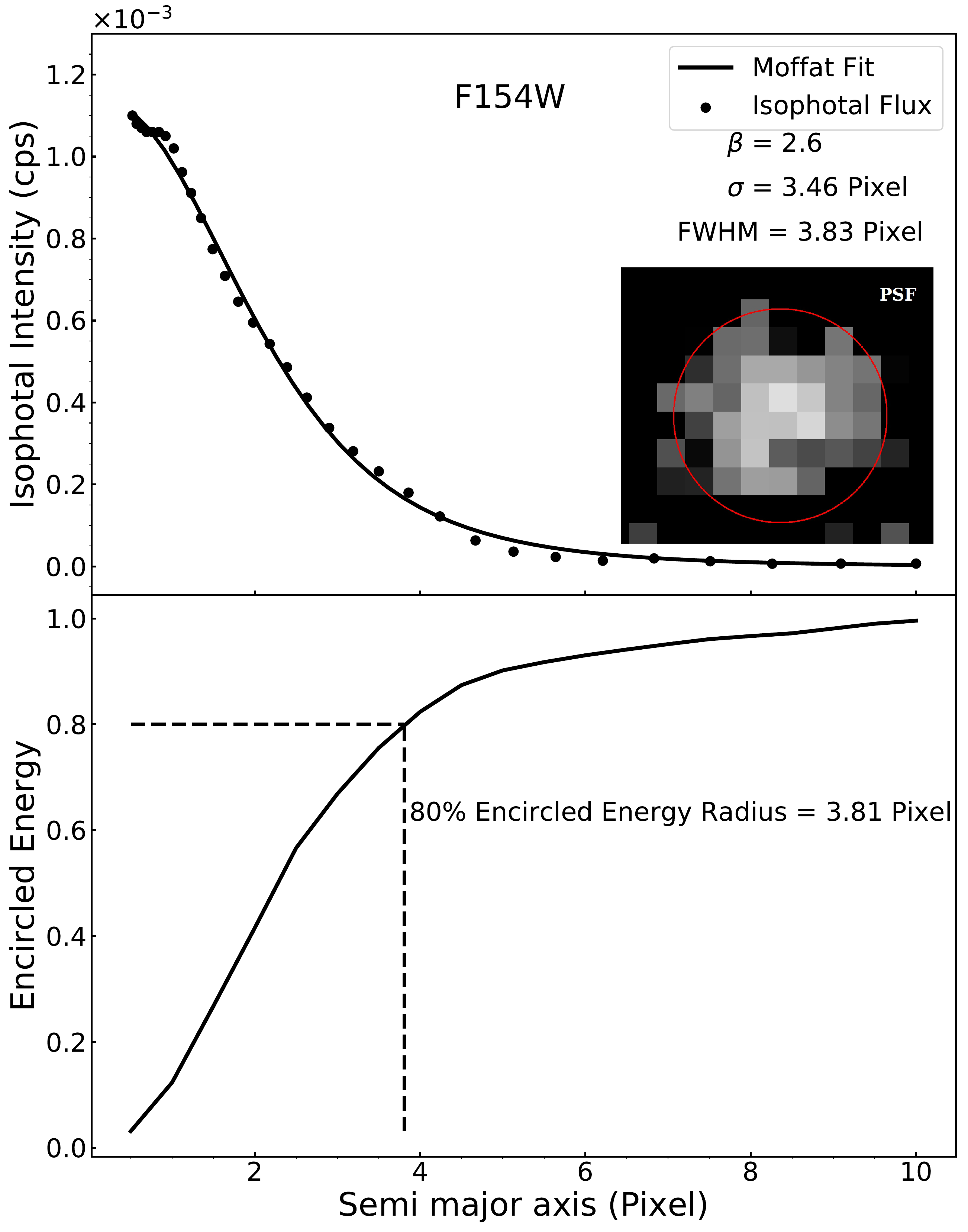}
    \includegraphics[width=0.43\linewidth]{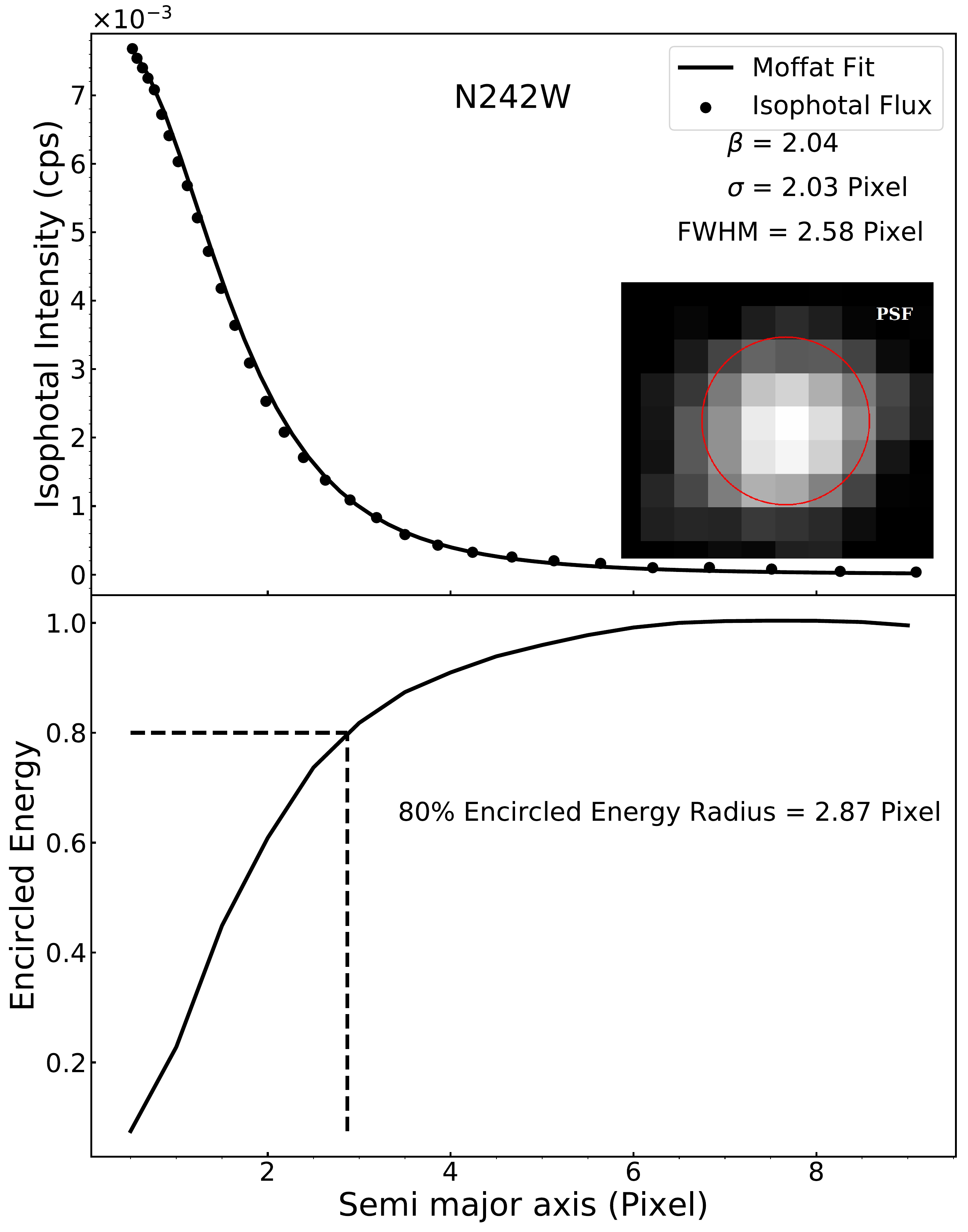}
     
    \caption{Top panels: Surface brightness distribution of the PSF is fitted with a circular Moffat function (see Eq.~\ref{eq:Moffat}) in FUV and NUV filters. FWHM refers to the full width at half maxima for the Moffat function. Inset images show the stacked PSF. Bottom panels: Encircled energy curve for FUV (left) and NUV (right). Vertical dashed lines in either case denote the radii containing 80\% encircled energy. one pixel=0$^{\prime\prime}$.417.}
    \label{fig:PSF}
\end{figure*}

\section{DATA REDUCTION AND ANALYSIS}
\label{sec:DRA}

The FUV and NUV observations were carried out by {\em AstroSat}/UVIT in the photon counting mode with a frame rate of $\sim 34 $~frames per second. This would accumulate about $\sim 45000 - 50000$ frames in a typical good dump-orbit. The orbit-wise dataset was processed using the official L2 pipeline in which we removed frames that are affected by the cosmic-ray shower and these were not included in the final science-ready images and the subsequent calculation of the photometry. This results in an average loss of $\sim 20\%$ data to science-ready images. The final science-ready image had a total exposure time of $t_{exp}=8600$~sec in FUV and $t_{exp}=7513$~sec NUV bands.
\par
Astrometric correction was performed using the {\em GALEX} FUV/NUV tiles and {\em SDSS} r-band image as references. We have used an IDL program that takes an input set of matched xpixel/ ypixel (from UVIT images) and RA/ DEC (from {\em GALEX} FUV/NUV and {\em SDSS} r band image) and perform a TANGENT-Plane astrometric plate solution similar to ccmap task of IRAF \citep{1993ASPC...52..173T}. The astrometric accuracy in NUV was found to be $\sim 0''.2$ while for FUV, the RMS was found to be $\sim 0''.24$, approximately half a pixel size. The photometric calibration was performed with a white dwarf star Hz4; the photometric zero-point for F154W band is 17.78 mag and 19.81 mag for N242W \citep{2017AJ....154..128T}. Once photometric calibration and astrometric correction are successfully applied, we run Source Extractor (SExtractor) software \citep{1996A&AS..117..393B} on the science-ready images to extract sources and estimate the background. We use following extraction parameters to detect sources in FUV and NUV images: {\tt DETECT\_THRESH} = 3- and 5$\sigma$ and {\tt DETECT\_MINAREA} = 16/ 9 pixels for FUV/ NUV filters depending on their angular resolution (See, section~\ref{sec:PSF_bkg}).

\subsection{Point Spread Function and Background estimation}
\label{sec:PSF_bkg}

We perform a robust calculation of full width half maxima (FWHM) measurements of the point spread function (hereafter, PSF) on the UVIT NUV and FUV science-ready images. We start with stacking a few unsaturated, isolated point sources of varying magnitudes to get an unbiased profile of a point source with a high signal-to-noise ratio (SNR) (See, inset image Figure~\ref{fig:PSF}). The isolated sources were selected from our UVIT FUV/NUV 3$\sigma$ catalog with an additional criteria of CLASS\_STAR (discussed in section~\ref{sec:CP}) $\geq$ $0.9$. Also, we visually examine the sources to check their symmetry around its centroid. In this work, we have adopted two methods: firstly, isophotal ellipse fitting is performed on the stacked images using IRAF STSDAS package\footnote{STSDAS is product of the Space Telescope Science Institute, which is operated by AURA for NASA.}. We fit the one-dimensional circular Moffat function over the surface brightness distribution $I(r)$ as a function of radius to obtain the parameters required for calculating FWHM \citep{1969A&A.....3..455M}. The Moffat function used to model the PSF is given by 
\begin{equation}
 I(r) = I_0 \left[1 + \bigg(\frac{ r } { \alpha }\bigg)^2  \right]^{-\beta}
\label{eq:Moffat}
\end{equation}
 
\noindent with FWHM = $ 2\alpha \sqrt{2^\frac{1}{\beta}-1}$. Here, $I_{0}$ is the central surface brightness and $\beta$ and $\alpha$ are the free parameters. ${\beta}$ is the seeing parameter that determines the spread of Moffat function. Here, we use Moffat function for simplicity as Gaussian function alone doesn't accurately fit the wings of a stellar profile \citep{2001MNRAS.328..977T} but see \cite{Sahaetal2021} for the wing modelling in F154W band. The Gaussian function is a limiting case of Moffat function (${\beta} \rightarrow {\infty}$). The second method for determining PSF involves creating encircled energy (EE) curve as a function of radius for the same stacked source. As per our calculation, the radius corresponding to a circular area enclosing 80\% of the total normalised energy of the stacked stellar profile came out to be close to PSF FWHM obtained using the first method. The results from both methods are in a good agreement as shown in Figure~\ref{fig:PSF}. The resulting values of the PSF FWHM for F154W and N242W filters from the fitting are 3.83 pixels (1$^{\prime\prime}$.59) and 2.58 pixels (1$^{\prime\prime}$.08).

\par
\begin{figure*}
    \includegraphics[width=0.45\linewidth]{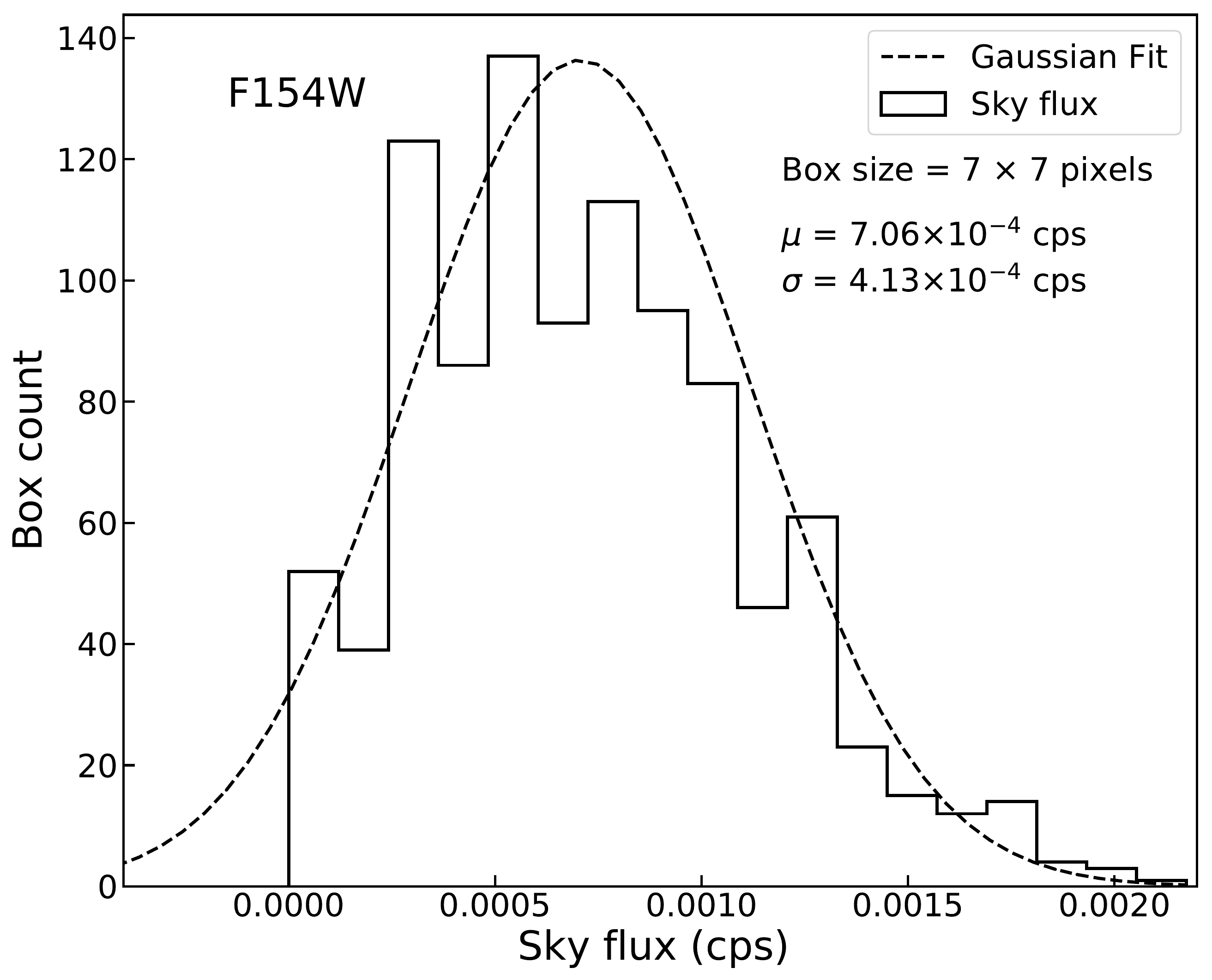}
     \includegraphics[width=0.45\linewidth]{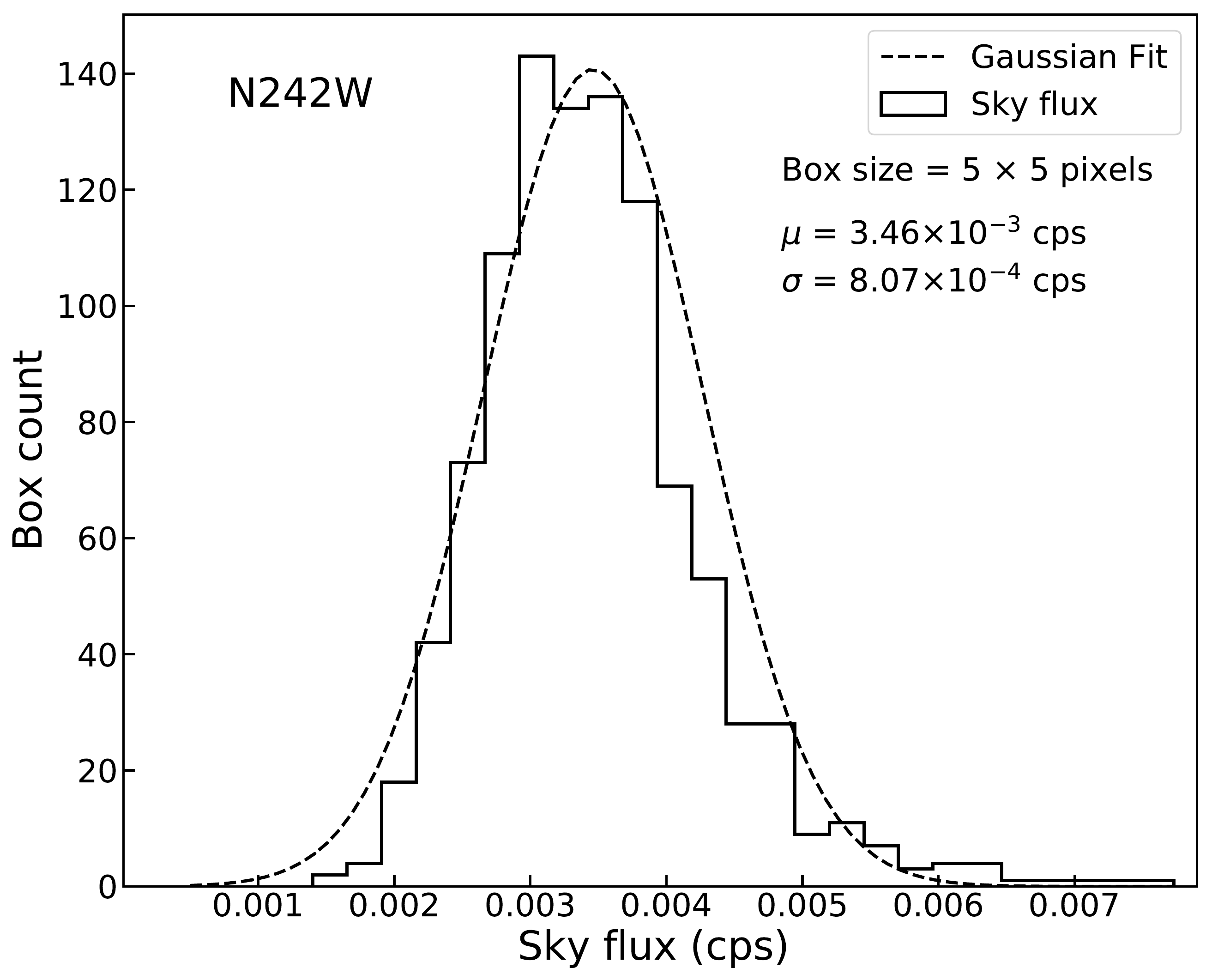}
     \caption{The figure comprises of background flux histograms for UVIT FUV/NUV filters. Background flux per pixel is calculated by fitting a single peak Gaussian function over the integrated flux distribution of 1000 boxes spread across the entire image.}
   \label{fig:sky}
\end{figure*}

We perform the background subtraction over the science-ready images in both filters for accurate flux estimation from the sources. For this, we run SExtractor \citep{1996A&AS..117..393B} with a detection threshold = 2$\sigma$ on both images. Subsequently, we mask all sources at and above 2${\sigma}$ from the entire FUV and NUV images. Thereafter, we measure integrated flux due to background by randomly placing multiple square boxes ($\approx$ 1000) over various parts of the masked image avoiding the source locations. The size of boxes (5${\times}$5 pixels for NUV/ 7${\times}$7 pixels for FUV) were chosen such that the area enclosed within the boxes were close to the area bounded by a PSF-size point source. Figure~\ref{fig:sky} shows the background flux histograms for the FUV and NUV images. These histograms are fitted with a Gaussian function with a mean ($\mu$) and standard deviation ($\sigma$). From the fitting, we obtain a sky surface brightness of 27.99 mag arcsec$^{-2}$ and 27.55 mag arcsec$^{-2}$ in the FUV and NUV observations, respectively. Measured mean background flux per pixel was subtracted prior to photometry. 

\par
\begin{figure*}
    \includegraphics[width=0.45\linewidth]{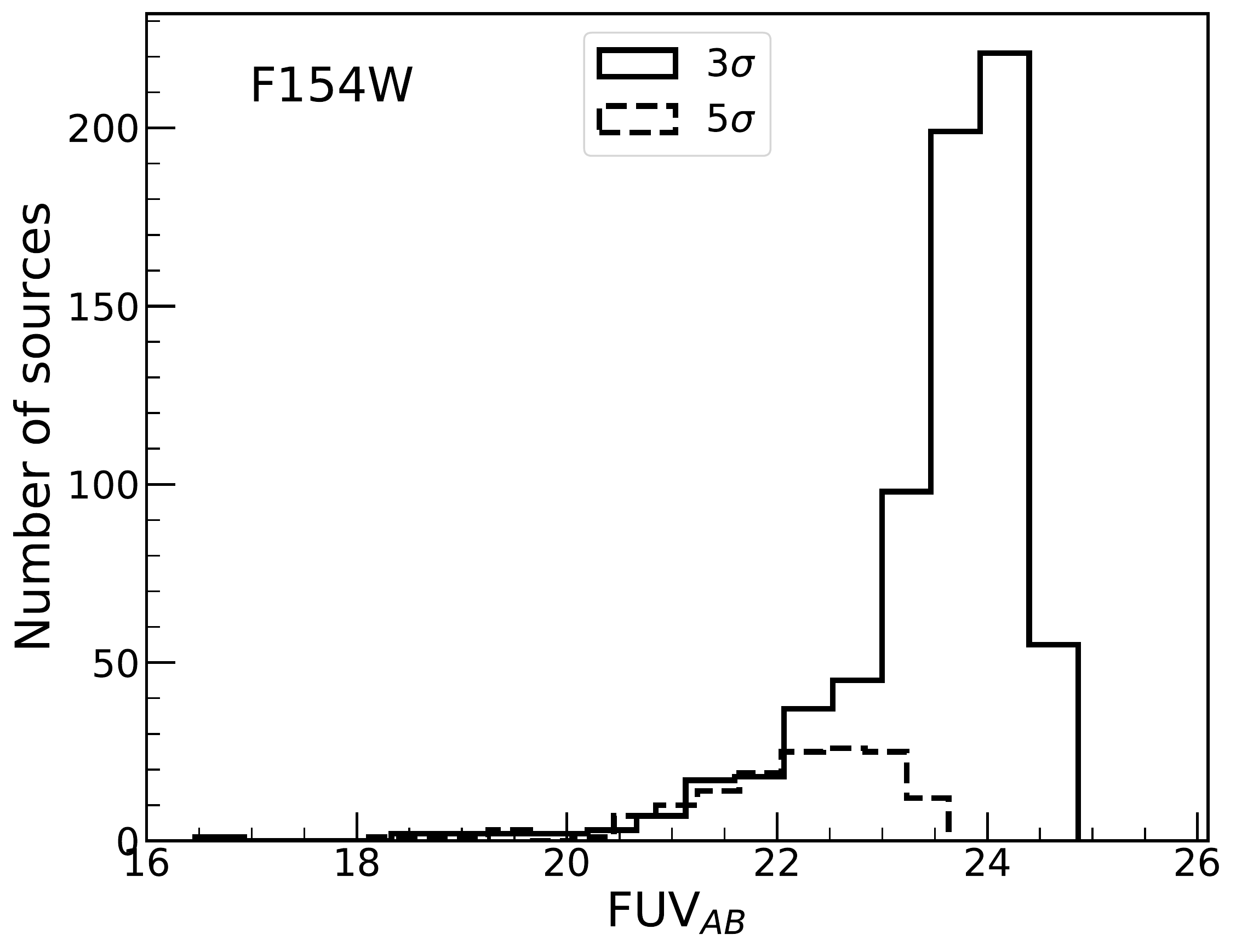}
    \includegraphics[width=0.45\linewidth]{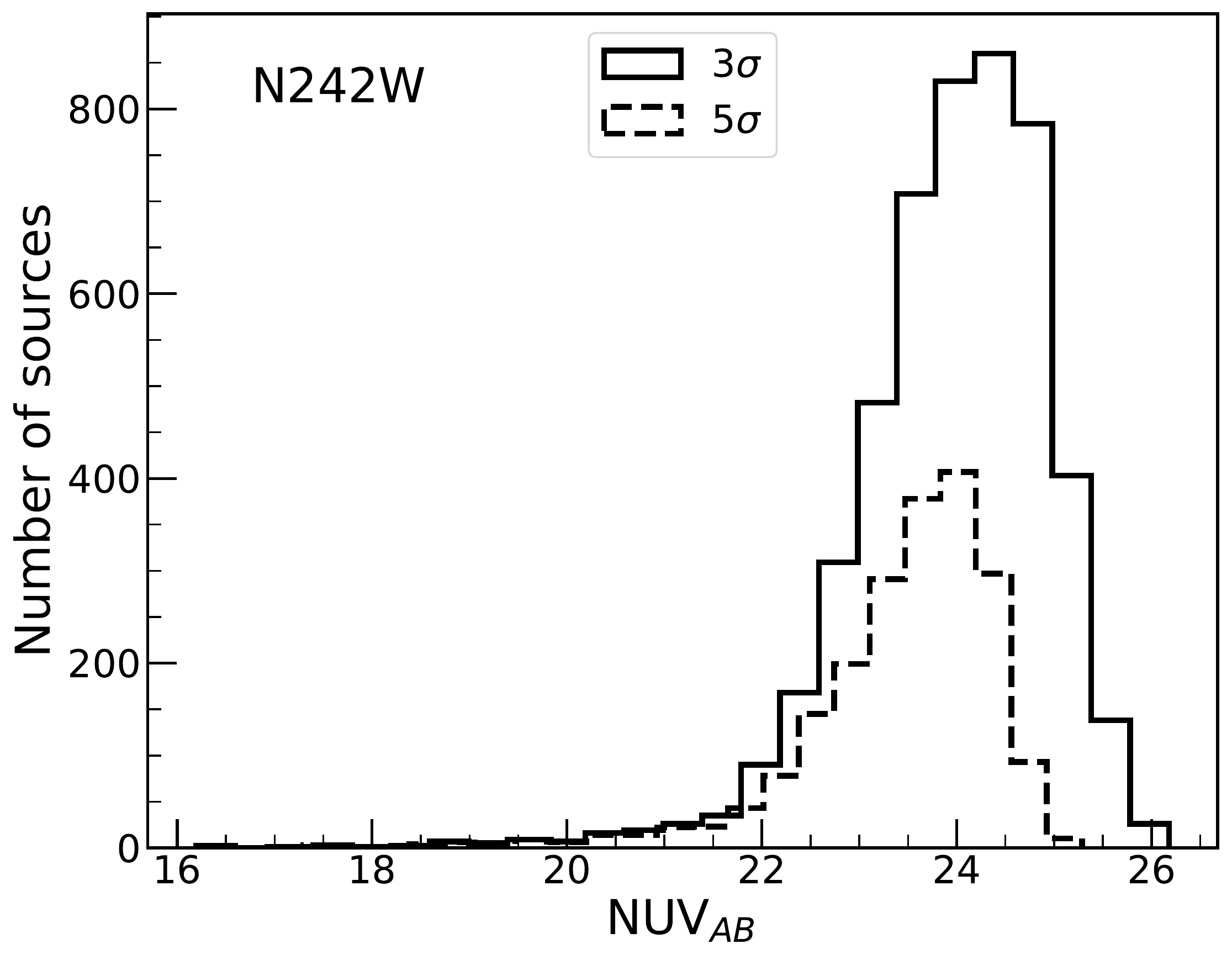} 
    \caption{The magnitude histograms for 3$\sigma$ and 5$\sigma$ FUV/NUV sources extracted by SExtractor \citep{1996A&AS..117..393B} and are not extinction corrected. Typical foreground dust extinction values for this FoV are $A_\mathrm{F154W}=0.24$ mag and $A_\mathrm{242W}=0.22$ mag. }
    \label{fig:comp_hist}
\end{figure*}

We follow Kron photometric technique \citep{1980ApJS...43..305K} in this work. The Kron apertures are elliptical or circular depending on the intensity distribution of the source, and the apertures capture 80-90\% of the total flux radiated by a source. We calculate the point source detection limit using the estimated background noise ($\sigma_\mathrm{sky}$), number of pixels in a circular aperture (\textit{r}), and desired detection threshold. The detection limits are calculated at 3${\sigma}$ and 5${\sigma}$ threshold considering aperture radius {\em r =} 4 pixels ($\sim$1$^{\prime\prime}$.6) and 3 pixels ($\sim $1$^{\prime\prime}$.2) for FUV and NUV images, respectively. We use $\sigma_{sky}$ corresponding to our FUV/NUV sky histograms (Figure~\ref{fig:sky}) $\approx$ $8.4$ $\times$ $10^{-6}$ /$3.23$ $\times$ $10^{-5}$ count s$^{-1}$ (cps). Note that these values are about a factor of $\approx$ 3 times lowered than their counterparts obtained using the SExtractor (FUV/NUV $\approx$ $2.2$ $\times$ $10^{-5}$ cps/$9.5$ $\times$ $10^{-5}$ cps). Nevertheless, the mean background obtained from both the methods are in good agreement. Using our $\sigma_{sky}$ from the histograms, the 3${\sigma}$ point source detection limit for FUV/NUV observations are found to be 25.02/26.22 mag. Similarly, our FUV/NUV survey reaches a detection limit of 24.46/25.66 mag at 5$\sigma$. Although we realize that the actual 3- and 5$\sigma$ detection limits depend on aperture size or the number of pixels within; shape of the object (point-like or extended), we consider this to be a good approximation of the actual UVIT detection limits. Figure~\ref{fig:comp_hist} shows magnitude histogram for all the extracted sources from FUV/NUV observation. From the magnitude histogram, the limiting magnitude for FUV/NUV observations at and above 3${\sigma}$ $\approx$ 24.0/25.0 mag. Whereas for 5${\sigma}$ threshold, the limiting magnitude for FUV/NUV $\approx$ 23.0/24.0 mag. Typical uncertainty in FUV/NUV magnitudes within the limit magnitudes is 0.3/0.2 mag. 

\par 
UV fluxes are highly susceptible to both internal and Galactic dust attenuation. \citet{1998ApJ...500..525S} full-sky 100 $\mu$m map gives us Galactic dust reddening $E(B - V)$ along a given line of sight. The values of extinction parameter $A_{F154W}/E(B-V) = 8.104$, $A_{N242W}/E(B-V) = 7.746$ were calculated using \cite{1989ApJ...345..245C} extinction law. We correct {\em SDSS} \textit{u}, \textit{g}, \textit{r}, \textit{i}, \textit{z} and {\em 2MASS} \textit{J}, \textit{H}, and \textit{Ks}-band fluxes for Galactic extinction in a similar manner. We use the method given by \citet{2012MNRAS.419.1727C} for redshift K-correction (K$_\mathrm{z}$) of the catalogue fluxes to $z = 0$. Consequently, we calculate the absolute magnitude (M$_\mathrm{\lambda}$) of a galaxy at luminosity distance D$_\mathrm{L}$ in a given passband (denoted by its wavelength $\lambda$) using the following equation(~\ref{eq:absolute_magnitude}). 
 
 \begin{equation}
 M_\mathrm{\lambda} = m_\mathrm{\lambda} - 5(\log D_\mathrm{L} - 1.0) - A_\mathrm{\lambda} - K_\mathrm{z}   
 \label{eq:absolute_magnitude}
 \end{equation}

\par
The cross-matching of sources over several observational surveys is a non-trivial task as morphologies and resolution may change over various wavelengths and filters used. Particularly, in our case, PSF FWHM for UVIT, {\em SDSS} and {\em 2MASS} survey are $\approx$ 1$^{\prime\prime}$.5, 1$^{\prime\prime}$.3 and 2$^{\prime\prime}$.8, respectively. The minimum cross-matching radius that we use for matching UVIT to {\em SDSS} catalogs $\sim$ 1$^{\prime\prime}$.5 whereas to {\em 2MASS} catalog $\sim$ 2$^{\prime\prime}$.8. Such values for cross-matching radius were selected as our field is a void field and to a large extent non-crowded. We find no multiple matches for any source present in our UVIT 3$\sigma$ catalog on cross matching with other catalogs from different surveys. We employ {\em topcat}\footnote{http://www.starlink.ac.uk/topcat/} software for the purpose. While we did the cross-matching, we also visually examined the sources in the UVIT FoV.

\par
%=================================
\begin{figure*}
    \centering
    \includegraphics[width=0.45\linewidth]{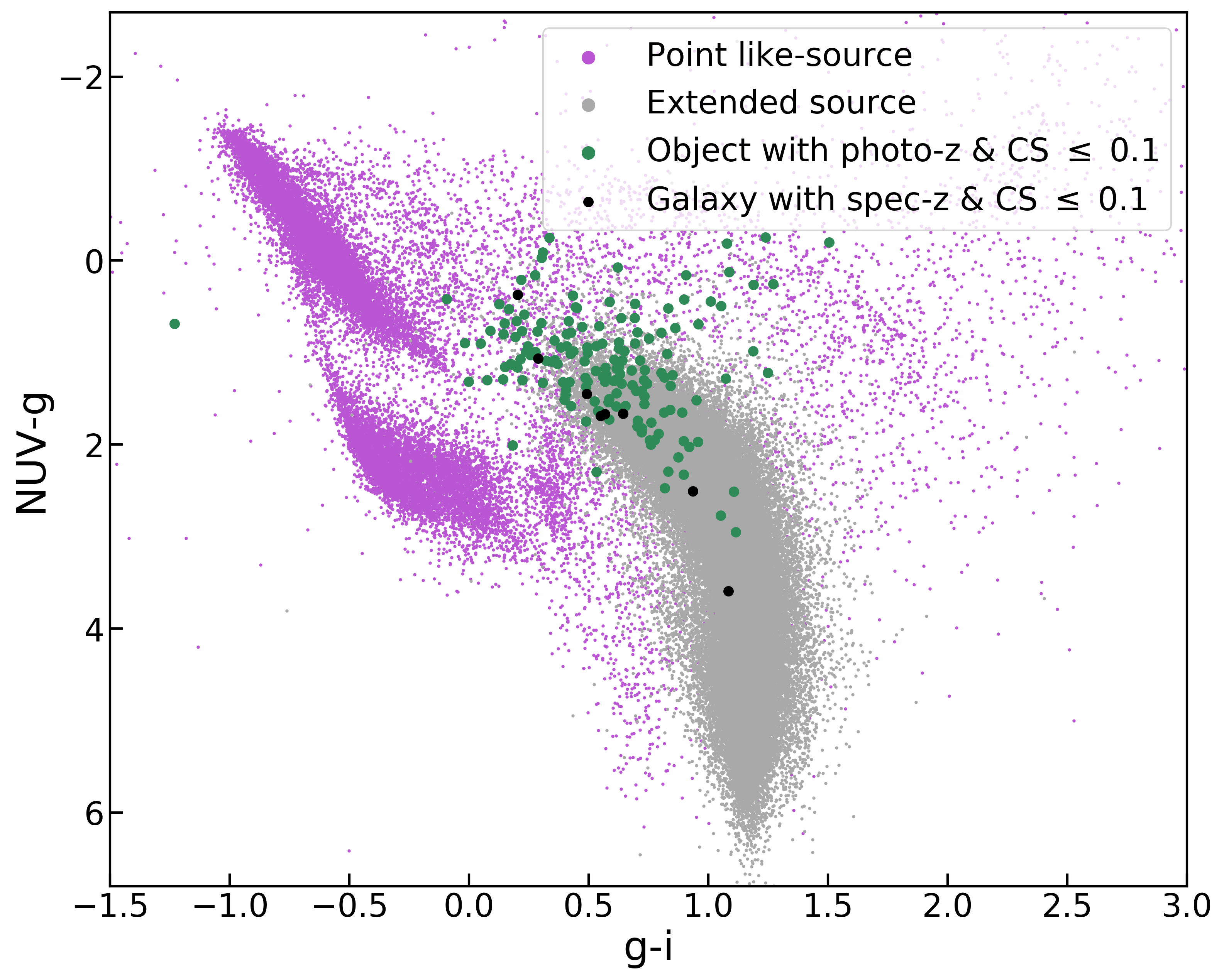}
    \includegraphics[width=0.45\linewidth]{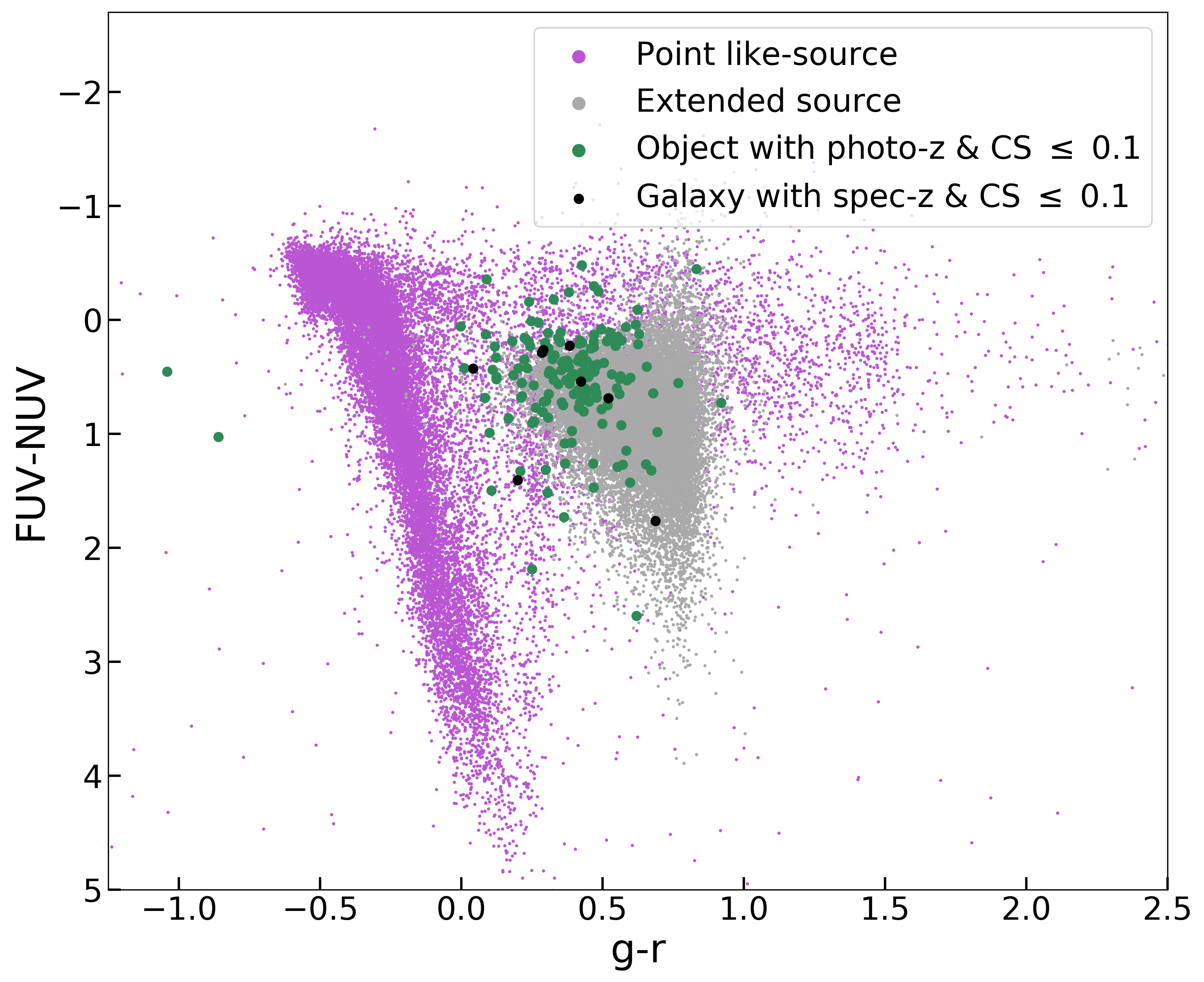} 
    \caption{NUV$-$g vs$.$ g$-$i and FUV$-$NUV vs$.$ g$-$r color-color diagrams used for $STAR/GALAXY$ classification of 159 sources with $z_\mathrm{{phot}}$ and CLASS\_STAR $\leq$ 0.1 from our UVIT 3$\sigma$ catalog are shown in left and right panels of the figure, respectively. The black markers represent galaxies from our UVIT 3$\sigma$ catalog with spectroscopic observation. The background point sources are taken from the overlap of archival {\em GALEX} AIS catalog and {\em SDSS} DR12 spectrocopic catalog whereas extended sources are from {\em GALEX-SDSS-WISE} Legacy catalog \citep{2016ApJS..227....2S}.}
    \label{fig:sgclassifcation}
\end{figure*}

\section{Catalog construction and Source classification}
\label{sec:CP}
The total number of sources extracted at 3$\sigma$ depth from FUV/NUV images are 709/4,931 and at 5$\sigma$ depth, 146/2,050 \footnote{UVIT 3$\sigma$ and 5$\sigma$ source catalog could be made available pertaining to an online request. These catalogs are going to be published online soon.}. As we are probing void galaxies in our FoV, we segregate the extended objects from the rest of the sources. At a rudimentary level, we use CLASS\_STAR (hereafter, CS) parameter given by SExtractor. The parameter gives a probabilistic value between zero (= $GALAXY$) to one (= $STAR$). In the current work, CS corresponding to {\em SDSS} r-band image is considered for the analysis where an object with CS $ \leq $ 0.1 is regarded as an extended source. As a result, we identify 184 and 74 extended sources with 3$\sigma$ and 5$\sigma$ detection, respectively. For all these extended sources, we have FUV and NUV observations. We are exclusively interested in sources with measured redshifts (spectroscopic/photometric). Therefore, on matching the extended sources (CS $\leq$ 0.1) present in UVIT FUV 3$\sigma$ catalog with {\em SDSS} spectrocopic/photometric catalog taken from {\em SDSS} DR12, we split the 184 extended sources in the following categories: 

\begin{itemize}
    \item Objects with photo-$z$ only = 159
    \item Objects with spec-$z$ = 8
    \item Objects absent in either of the catalogs (spectrocopic/photometric) = 17 
 \end{itemize}

Visual examination of the sources which were not a part of any of the two catalogs reveals that some of them are bright and saturated stars, part of an extended source, or faint sources ($u \sim$ 23 mag). Hence, we reject the objects in this category for any further analysis. All 8 extended sources with spec-$z$ were already classified as $GALAXY$ by $SDSS$. In addition, we find a pea shaped/compact object having spec-$z$ in our FoV with CS $>$ 0.1, classified as $GALAXY$ by $SDSS$. The 159 extended sources present in UVIT 3$\sigma$ catalog with photo-$z$ are subjected to rigorous classification methodology.     
\par
For the identification of 159 objects as galaxies, we use color-color diagrams described in \citet{2007ApJS..173..659B} that classifies photometrically selected sources into various categories of astrophysical origin. This method involves broad band photometric data combining seven passbands from {\em GALEX} (FUV/NUV) to {\em SDSS} (\textit{u, g, r, i, z}). We use NUV$-$g vs$.$ g$-$i and FUV$-$NUV vs$.$ g$-$r color-color diagrams as extended sources (mostly galaxies) are well separated from point-like (Star/QSO) sources on these color-color planes (Figure~\ref{fig:sgclassifcation}). For this purpose, we curate a catalog of point sources combining archival {\em GALEX} all-sky imaging survey (AIS) observations with archival {\em SDSS DR12} spectroscopic catalog. The point sources present in the catalog are spectroscopically classified as $STAR$ by {\em SDSS}. On the other hand, the extended source catalog comprising of galaxies are taken from {\em GALEX-SDSS-WISE} Legacy catalog \citep{2016ApJS..227....2S} (Hereafter, $Salim\text{ }Catalog$). We plot our 159 objects along with the sources present in the point source and extended catalogs on the color-color diagrams to observe that our UVIT sources mostly overlap the region corresponding to extended sources in both color-color diagrams and therefore, we consider these objects as extended sources(or galaxies) for further analysis in the work. 
\par
On matching {\em SDSS} spectroscopic catalog with our UVIT $3\sigma$ catalog, we confirm two galaxy members of the Bootes Void. The photometric redshifts given by {\em SDSS} ($z^\prime$) have comparable (of the same order) errors ($z_\mathrm{{err}}^\prime$) associated with them. Hitherto, $z^{\prime}$ for none of the galaxies fall within the redshift of the Bootes Void i$.$e$.$, 0.04 $\leq z \leq$ 0.06 \citep{1987ApJ...314..493K,1981ApJ...248L..57K}. However, nine galaxies are identified out of 159 photometrically selected extended sources such that $z{^\prime \pm z{_\mathrm{{err}}^\prime}}$ falls within the void's redshift range. In the following section, we determine our own set of photometric redshifts for galaxies with $z^\prime$ by modelling their broadband spectral energy distribution to select the void galaxy candidates with increased certainty.

\subsection{Determination of Photometric Redshifts and assigning candidature to the Bootes Void}
\label{sec:pz}
\begin{deluxetable*}{ccccccccccc}
% \tablenum{1}
\tablecaption{Void candidature determination based on cumulative distribution function for EAZY redshifts \label{tab:eazy_out}}
% \tablewidth{0pt}
\tablehead{
\colhead{G No.}&\colhead{RA}&\colhead{DEC}&\colhead{FUV$_\mathrm{GALEX}$}&\colhead{NUV$_\mathrm{GALEX}$}&\colhead{FUV$_\mathrm{UVIT}$}&\colhead{NUV$_\mathrm{UVIT}$}&\colhead{$z^\prime$}&\colhead{$z$}&\colhead{n}&\colhead{P($z$)}}
\colnumbers
\startdata
    G1&14:07:25.63&48:50:43.4&20.32$\pm$0.14&19.94$\pm$ 0.09&20.29$\pm$0.05& 19.80$\pm$0.01&0.029$\pm$0.23&0.055$\pm$0.012&1&0.54 \\
    G2&14:07:39.55&48:57:33.1&-&-&23.31$\pm$0.22&23.07$\pm$0.07&0.449$\pm$0.154&0.057$\pm$0.064&2&0.12 \\
    G3&14:08:43.44&48:54:10.8&-&22.55$\pm$0.41&22.27$\pm$0.14&22.47$\pm$0.05&0.102$\pm$0.047&0.043$\pm$0.011&1&0.56\\
    G4&14:07:53.33&49:01:13.1&21.70$\pm$0.32&21.74$\pm$0.25&21.52$\pm$0.09&21.91$\pm$0.03&0.177$\pm$0.094&0.057$\pm$0.003&1&0.77 \\
    G5&14:09:09.34&49:08:49.2&-&21.97$\pm$0.28&22.20$\pm$0.13&22.05$\pm$0.04&0.253$\pm$0.182&0.053$\pm$0.055&2&0.14\\
    G6&14:07:40.08&49:05:10.5&21.10$\pm$0.19&20.73$\pm$0.14&20.87$\pm$0.07&20.58$\pm$0.02&0.078$\pm$0.023&0.055$\pm$0.014&1&0.50\\
\enddata
\tablecomments{Col ID: (1) Galaxy No., (2) RA J2000, (3) DEC J2000, (4) GALEX FUV mag, (5) GALEX NUV mag, (6) UVIT FUV mag, (7) UVIT NUV mag, (8) photometric redshift given by SDSS,  (9) photometric redshift calculated using EAZY, (10) n = $\frac{| z-{\mu_z}|}{\sigma_z}$, (11) cumulative probability of the galaxy at redshift $z$ to be located in the redshift range of the void P($0.04 \leq z \leq 0.06$). Magnitudes in the table are not corrected for Galactic extinction.}
\end{deluxetable*}

We determine the photometric redshifts of all 159 photometrically selected galaxies using the photo-$z$ code called EAZY \citep{brammer2008}. In the process, the photometric fluxes of seven broadband filters were utilized namely, UVIT {(\em FUV, NUV)} and {\em SDSS (u, g, r, i, z)}. EAZY provides us ${\chi}^2$ minimized redshift for the best fit linear combination of all galaxy templates. We use six standard galaxy templates in our calculations and derive photometric redshifts ($z_\mathrm{phot}$) for all galaxies. The procedure is inclusive of the wavelength dependent template error function. 

\begin{figure*}
    \centering
    \includegraphics[width=0.48\linewidth]{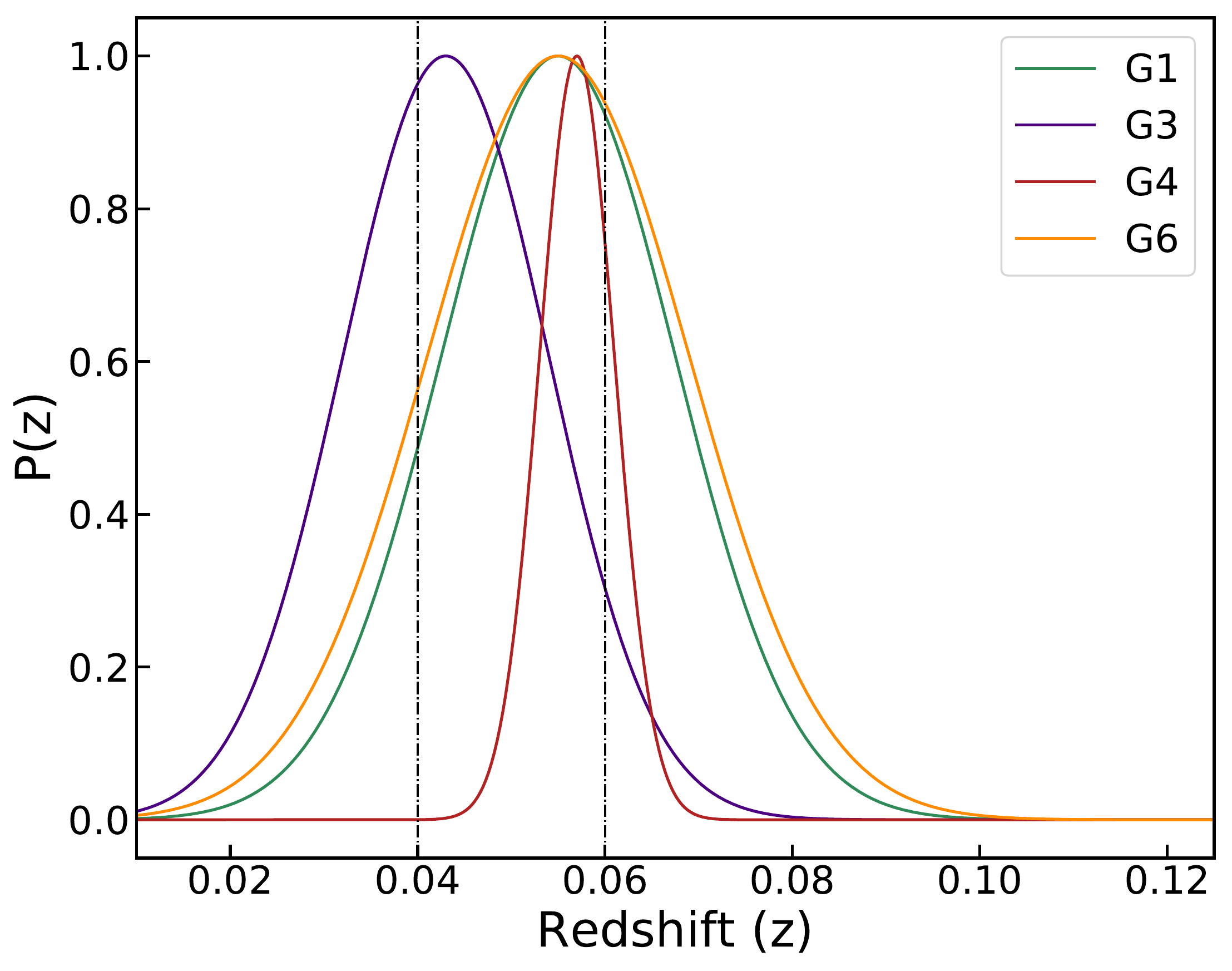}
    \includegraphics[width=0.48\linewidth]{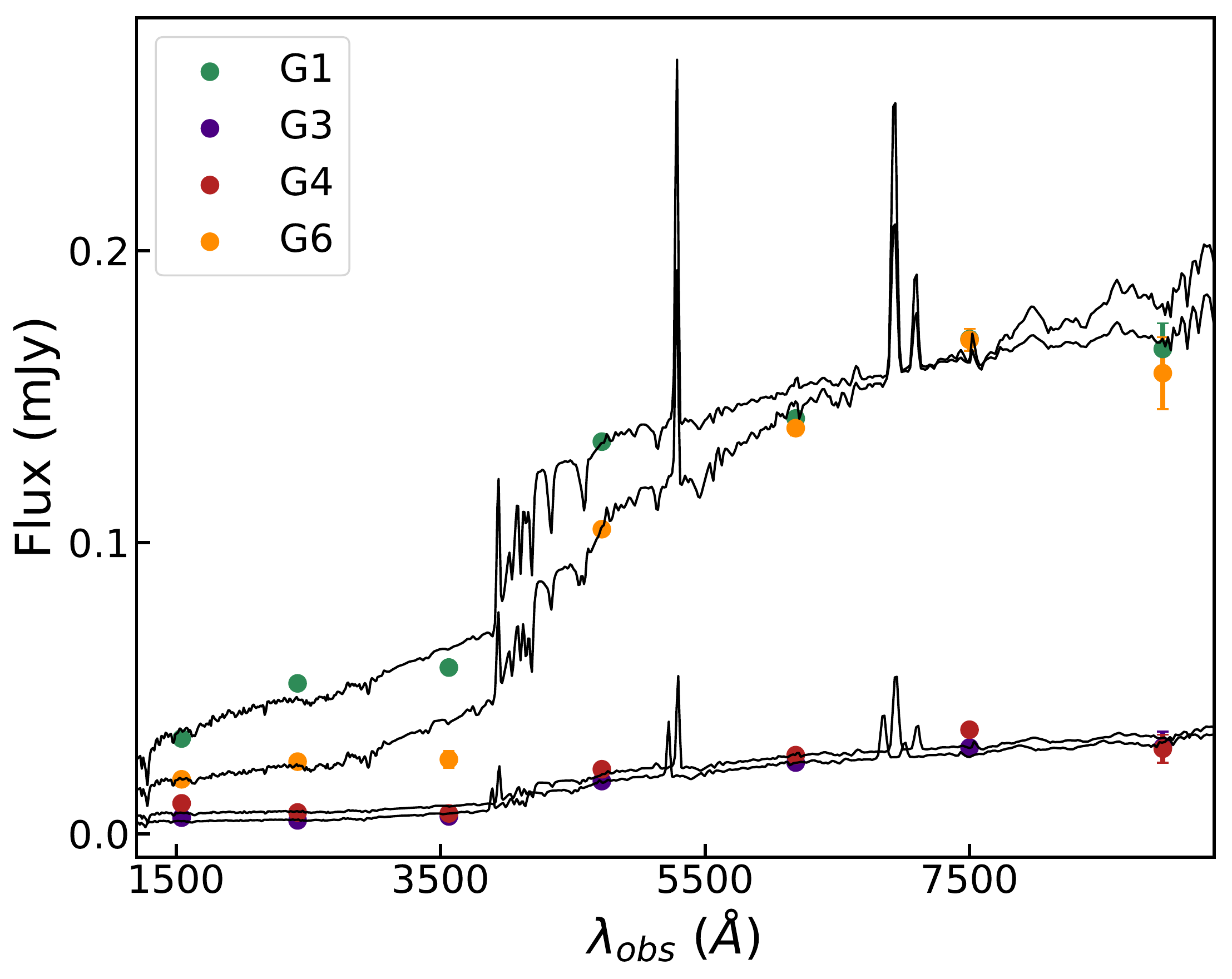} 
    \caption{Left Panel: The probability density function P(z) as a function of redshift color-coded for four void galaxies with $z_{\rm phot}$ based on our EAZY redshift estimation with P(z) $\geq$ 0.5. Redshift extent of the Bootes Void marked with black dotted lines. The galaxies are numbered according to Table~\ref{tab:eazy_out}. Right Panel: EAZY best fit spectral energy distributions with observed fluxes for the four void galaxies. The galaxies are color-coded and numbered same as the Left panel. }
    \label{fig:EAZY}
\end{figure*}

In order to find the quality of our fit, we deduce photometric redshifts for the two void galaxies with $z_\mathrm{spec}$ in a similar manner with an equal number of broadband filter magnitudes and calculate $\Delta z = |z_\mathrm{phot}-z_\mathrm{spec}|/(1+z_\mathrm{spec})$. The quantity ${\Delta z}$ for the two void galaxies averages to $\approx$ 0.01 wherein conventionally one discards $z_\mathrm{phot}$ with $\Delta z$ $>$ 0.1 \citep{skelton2014}. The result indicates fair agreement between the EAZY photo-z  and SDSS spectroscopic redshifts. EAZY also provides us with 1-, 2- and 3$\sigma$ confidence intervals computed from posterior probability distribution for each galaxy. The mean redshift ($\mu_z$) and sigma ($\sigma_z$) for the probability distributions were calculated post assuming these distributions as a single peak Gaussian function. Nearly, all $z_\mathrm{phot}$ fall within 1$\sigma$ confidence limits of the posterior probability distribution (see $n$ in Table~\ref{tab:eazy_out}) and the uncertainty in $z_\mathrm{phot}$ were given by $|z_\mathrm{phot}-{\mu_z}|$.

EAZY photometric redshift of six galaxies lie in the redshift range of the Bootes Void. The cumulative probabilities for all six galaxies to exist inside the void corresponding to their EAZY redshift (P($0.04 \leq z \leq 0.06$)) and the output redshifts are given in Table~\ref{tab:eazy_out}. Evidently, P($z$) gives a clear depiction of the void candidature among the six galaxies. With a cut of P($z$) $\gtrsim$ 0.5, we further narrow down our potential void candidates sample to four galaxies (G1, G3, G4, and G6) with $z_\mathrm{phot}$ which we intend to study further in the work. Furthermore, the probability distribution function P($z$) as a function of $z$, and the model spectral energy distributions and observed fluxes for the four galaxies are shown in the left and right panel of Figure~\ref{fig:EAZY}, respectively.

\par
\begin{figure}
%    \centering
	\includegraphics[height =9cm,width=9.0cm]{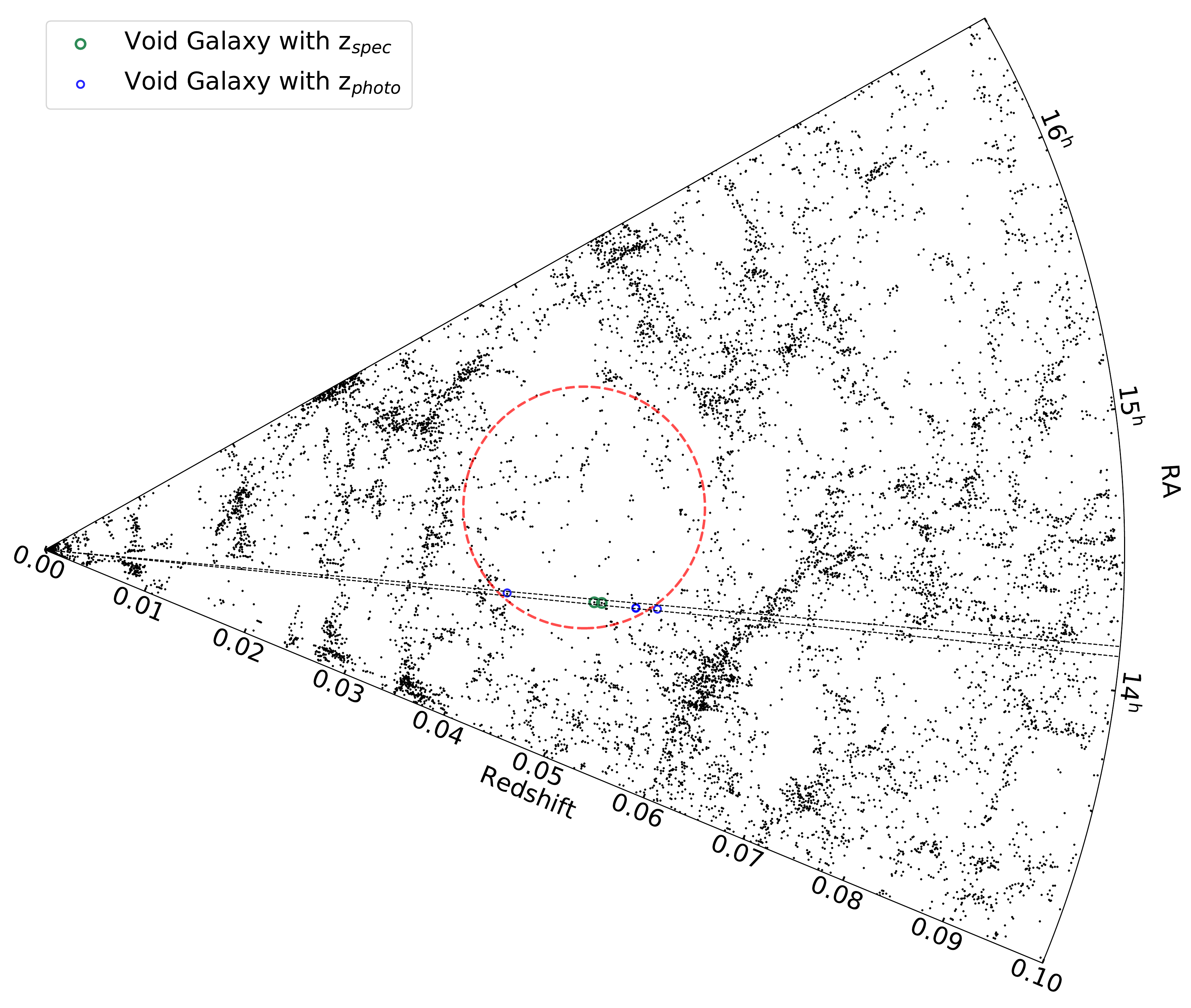}
    \caption{The cone diagram shows the redshift distribution of galaxies in the direction of Bootes Void. The black dots in the wedge diagram are sources taken from {\em SDSS} DR12 spectroscopic catalog and the pair of black dashed lines encloses the solid angle subtended by our UVIT pointing. The red open circle marks the boundary of the Bootes Void; green open circles are the confirmed void galaxy members (with $z_\mathrm{{spec}}$); blue open circles represents void galaxies with $z_\mathrm{{phot}}$. The likelihood for each galaxy with $z_\mathrm{{phot}}$ to be located in the void is higher than 50\% (see Table~\ref{tab:eazy_out}). 
    %Follow-up spectroscopic observation is required to confirm void candidature of void galaxies with $z_\mathrm{phot}$
    }
    \label{fig:cone}
\end{figure}

With this work, we report four newly detected void galaxies with $z_\mathrm{phot}$ (G1, G3, G4 and G6) along with two previously identified void galaxies with $z_\mathrm{spec}$. Figure~\ref{fig:cone} shows the redshift distribution of galaxies in the direction of the Bootes Void. The void is roughly considered as spherical in shape. The red circle marks a radius of $\sim$ 46 Mpc from the center and roughly denotes the boundary of the void. As described in \citet{Kirshner1983}, we assume the center of the void at $\alpha$ = 14$^h$ 48$^m$, $\delta$ = +47$^d$ and mean redshift of $\sim$ 0.05. Most of our UVIT detected void galaxies lie close to the boundary of the void. A follow up spectroscopic survey would be required to confirm void membership of the four galaxies with z$_\mathrm{phot}$. \par 

\section{UVIT Detections}
\label{sec:detections}

Prior to UVIT, {\em GALEX} has observed our FoV in both FUV and NUV passbands as part of its all-sky imaging survey (AIS). We compare the SNRs of the void galaxies with $z_\mathrm{phot}$ in {\em GALEX} and UVIT FUV observations to showcase the enhanced sensitivity of the UVIT deep imaging survey. We use the  following equation for our calculation of SNRs \citep{Sahaetal2020}. 

\begin{equation}
 SNR = \frac{F_\mathrm{g} t \epsilon}{\sqrt{F_\mathrm{g} t \epsilon + f_\mathrm{b} n_\mathrm{pix} t \epsilon} } ,    
 \label{eq:snr_eq}
\end{equation}

\noindent where $f_\mathrm{b}$ denotes the background noise per pixel (estimated from the final science-ready images) and $t$ is exposure time for UVIT FUV observation as mentioned in section~\ref{sec:DRA}. $F_\mathrm{g}$ denotes the total number of detected photons from the source alone within a given aperture containing $n_\mathrm{pix}$ pixels measured from the science-ready images. In other words, $F_\mathrm{g}$ denotes the total number of detected photons minus the number of background photons from the same aperture. For comparison with GALEX observation, we consider a circular aperture of $r$ $\approx$ 2$^{\prime\prime}$.5 at a fixed position corresponding to the RA/DEC of sources present in the {\em SDSS} catalog placed on {\em GALEX} and UVIT FUV images to estimate the total detected photons from the galaxies ($F_\mathrm{g}$). We have visually checked that there are no other sources within this circular aperture. The values of $f_\mathrm{b}$ and $t$ for the particular {\em GALEX} FUV tile are $3 {\times}$ 10$^{-4}$ cps and 205 sec, respectively. 
\par
\begin{figure}
    \centering
    \includegraphics[width=8.5cm]{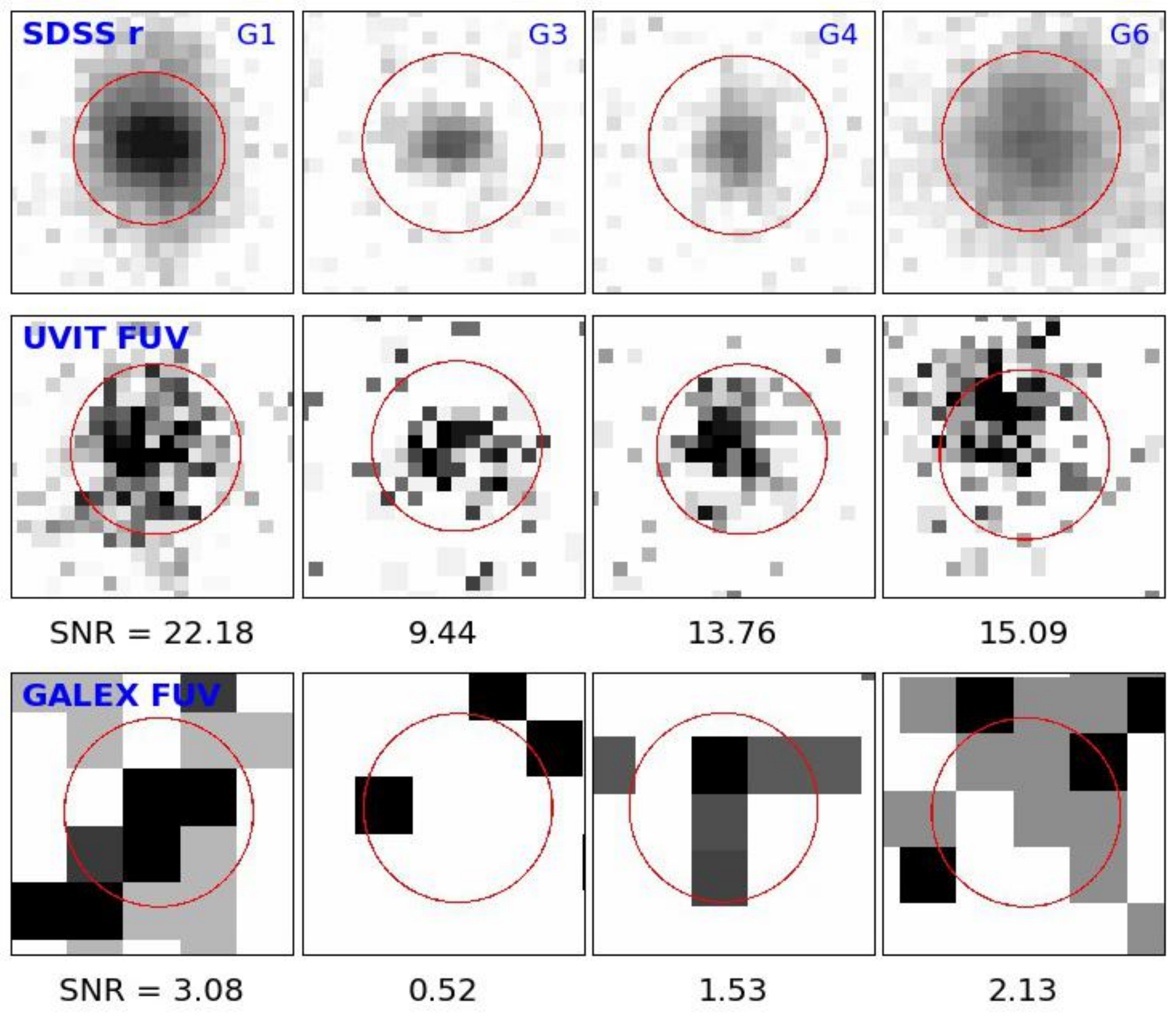}
    \caption{The figure shows the cutout images of four void galaxies with $z_\mathrm{phot}$ in three filters (top row:{\em SDSS} r, middle row: UVIT FUV and bottom row: {\em GALEX} FUV). The blue colored numbers in the top row are assigned to the galaxies according to Table~\ref{tab:eazy_out}. The center of circles inscribing the sources is as per the RA/DEC given in {\em SDSS} catalog. The radius of each circle shown in the figure is 2$^{\prime\prime}$.5. The signal-to-noise ratios (SNRs) for all sources estimated from UVIT and {\em GALEX} FUV observations are written beneath the middle and bottom row, respectively. }
    \label{fig:snr}
\end{figure}

 
Often, there is a small to moderate variation in the exposure time across the FoV due to the fact that edges receive less exposure than the center of the FoV. Therefore, we correct our SNRs for this effect by introducing a factor $\epsilon$ in Equation~\ref{eq:snr_eq}. We determine $\epsilon$ by taking the ratio of mean effective exposure across the image to the maximum exposure received at the center of the image. 
In the case of GALEX, we use a high resolution relative response map described in \citet{Morrisey07} corresponding to our FoV and evaluate $\epsilon$ $=$ 0.81. When we closely examine the UVIT exposure map generated from the L2 pipeline, we do not find such a gradual decrease in the effective exposure time as we move outwards at least upto 13$^{\prime}$ from the center. However, beyond 13$^{\prime}$ radius, the exposure time falls off sharply. On considering the entire FoV, we deduce $\epsilon$ $=$ 0.98 for UVIT. Interestingly, within 13$^{\prime}$ from the centre of the  FoV, we note that there can be a variation of $\simeq 2$\% in the exposure time over 8 pixels (exposure map appears to have a Moire pattern). This variation induces $\simeq 1$\% uncertainty in the reported SNRs, see Figure~\ref{fig:snr}.

\par
Our estimates of the SNRs are labeled in Figure~\ref{fig:snr} with the cutout images of the four void galaxies from {\em SDSS} r, UVIT FUV and {\em GALEX} FUV images. We observe that the UVIT SNRs are much higher than the {\em GALEX} SNRs. Only one galaxy (G1 in Figure~\ref{fig:snr}) cross the limiting SNR = 3 in {\em GALEX}. Based on our SNR calculation and photo-$z$ estimation, we find three new void  galaxies in our FUV observation i$.$e$.$, G3, G4 and G6.
\par
In addition, we refer to {\em GALEX} merged catalog for procuring FUV/NUV magnitudes with errors and for the size of Kron apertures used for photometry for all six void galaxy candidates \citep{Bianchi_2017}. Table~\ref{tab:eazy_out} lists GALEX and UVIT FUV/NUV magnitudes for all void galaxy candidates. Of these, three galaxies fainter than 22 mag in UVIT FUV observation are not detected in {\em GALEX} catalog (G2, G3 and G5). It is worth mentioning here that {\em GALEX} AIS reaches a typical 5$\sigma$ depth of $\sim$ 20 AB mag in FUV \citep{Bianchi_2017}. Moreover, on overlaying {\em GALEX} FUV Kron apertures on {\em SDSS} cutouts, we find a few nearby sources within the apertures in case of G1 and G4 making them unfit in the subsequent analysis. 

\par
\begin{figure}
\begin{flushleft}
	\includegraphics[height=8cm,width=9cm]{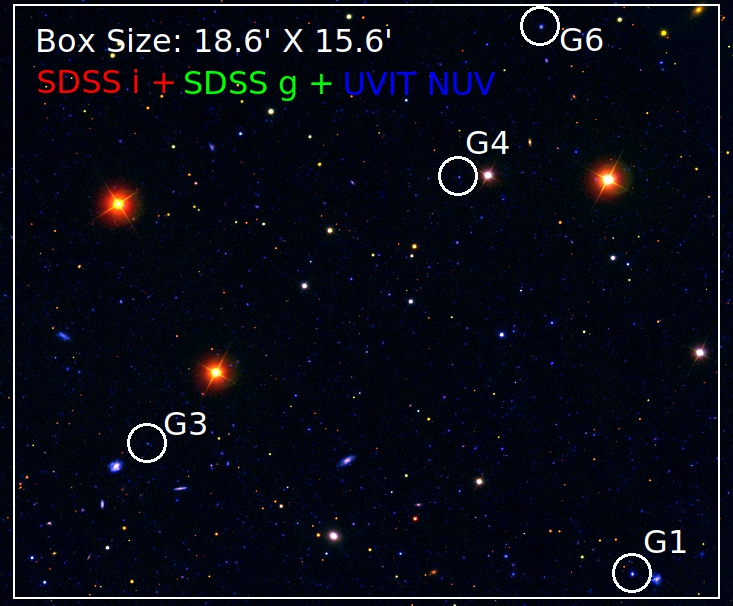}
    \caption{The color composite image (Red: {\em SDSS} i-filter, Green: {\em SDSS} g-filter, Blue: UVIT NUV filter) highlights the four void galaxies out of which three (G3, G4, and G6 in Figure~\ref{fig:snr}) are undetected in {\em GALEX} FUV observation. The serial numbers given to the galaxies are in reference to Table~\ref{tab:eazy_out}. 
    %The figure illustrates the importance of using deep imaging data provided by UVIT
    }
    \label{fig:i+g+nuv}
    \end{flushleft}
\end{figure}

Figure~\ref{fig:i+g+nuv} shows a color composite image of a portion of our FoV using {\em SDSS} \textit{i}, \textit{g} and UVIT NUV filters. We highlight all four void galaxies discussed earlier in this section. The result underlines the importance of using the UVIT deep observations over {\em GALEX} AIS data for this analysis.

\section{EXTINCTION IN THE UV CONTINUUM}
\label{sec:DE}
Internal dust present within a galaxy scatters and/or absorbs UV photons which makes it strenuous to estimate the absolute UV flux emitted from a galaxy. Several factors such as the geometry of a galaxy, amount of dust and its components affects the intensity of UV flux attenuation in a galaxy. There are various dust attenuation laws available for local and high redshift star-forming galaxies \citep{1994ApJ...429..582C,2000ApJ...533..682C,reddy2015mosdef}. Different methods could be used to solve for dust obscuration of UV photons which are based on two principles: Using slope (${\beta}$) of a power-law function ($f_{\lambda} = {\lambda}^{\beta}$) followed by UV continuum emission of galaxies over the wavelength range 1300 - 2600 \AA\ \citep{1994ApJ...429..582C}. The other method is based on total energy budget of a galaxy and it represents a combination of FUV and IR luminosities \citep{Hao_2011}.  
\par 
UV ${\beta}$ slope efficiently works as a diagnostic for internal dust attenuation \citep{beta_hz,M99} and far-IR luminosities are unavailable for our entire sample of void galaxies. Therefore, we use a method based on UV spectral slope ${\beta}$. The values of $\beta$ are calculated using the following relation \citep{Nordon_2012}:

\begin{equation}
{\beta} = -\frac{m({\lambda}_{1})-m({\lambda}_{2})}{2.5\log(\frac{\lambda_{1}}{\lambda_{2}})}-2
\label{eq:beta}
\end{equation}

\noindent Here, ${\lambda}_{1}$ and ${\lambda}_{2}$ are the effective wavelengths corresponding to UVIT FUV and UVIT NUV filters. The slope, thus, calculated can be used to find color excess $E(B-V)$ using the following relation~\ref{eq:ebv} \citep{Reddy_2018}:
\begin{equation}
{\beta} = - 2.616 + 4.684 E ( B - V )
\label{eq:ebv}
\end{equation}

\noindent The relation is derived using Calzetti $+$ 00 dust curve \citep{2000ApJ...533..682C} on BPASS galaxy model \citep{Reddy_2018}. The dust attenuation law established by \citet{2000ApJ...533..682C} is used to find the value of k(${\lambda}$) for F154W filter of UVIT. The extinction relation, A$_\mathrm{\lambda} = k({\lambda})E(B-V)$ (where $k(1541\,  \angstrom) = 10.18$) gives us total extinction in the FUV filter. The resultant A$_\mathrm{FUV}$ vs$.$ ${\beta}$ curve obtained by the above discussed method is less steeper than \citet{M99} curve. The slope $\beta$ provides us rough estimates of the ongoing star formation and internal dust obscuration of a galaxy \citep{Reddy_2018}. Lesser negative values of $\beta$ symbolises either the abundance of old stellar type or high internal dust concentration within a galaxy. ${\beta}$ for our sample ranges from $-$2.72 to $-$0.60 with median $\approx$ $-$1.35 indicating active ongoing star formation with low to moderate internal dust obscuration \citep{beta_slope}.
\par 
Henceforth, the intrinsic FUV luminosities of galaxies are used to calculate the FUV SFRs as described in the next section~\ref{sec:sfr}. In the following part of the work, unless mentioned otherwise, all colors and absolute magnitudes are corrected for Galactic extinction only, while the SFRs reported are corrected also for internal extinction.

\section{Stellar mass estimation}
\label{sec:Mstar}
Stellar masses (M$_{*}$) of galaxies are widely considered as one of the fundamental parameters that drive galaxy evolution over cosmic time. Not only galaxy evolution, unbiased, robust estimate of stellar masses can play a crucial role to constrain models of galaxy formation as well \citep{Salmonetal2015}. We estimate stellar masses of the void galaxies and the remaining non-void galaxies upto z $\leq$ 0.1 present in the FoV using two methods.
\par
In our first method, we perform broad-band (from AstroSat/UVIT far-UV to {\em SDSS} z-band) SED modelling using Code Investigating GALaxy Emission (CIGALE) \citep{cigale}; similar to previous section~\ref{sec:pz}. Our SED modelling proceeds with standard assumptions for the star formation histories (SFH), initial mass function (IMF), dust attenuation, etc$.$. We adopt a double exponential function for the SFH with $SFR(t)\text{ }{\propto}\text{ }exp(-t/{\tau})$ forming bulk of the stellar mass and another exponential function to accommodate the recent burst of star formation. In the previous expression, $t$ is the time since onset of star formation and ${\tau}$ is e-folding timescale. The young and old stellar populations are separated by 10 Myr. The intrinsic stellar population in the galaxy is modelled with a \citet{BC03} stellar population library. We choose \citet{Salpeter} IMF with a range of masses varying from 0.1 - 100 M$_{\odot}$ for determining the intrinsic population. The metallicity for each galaxy was given as a free parameter (to chose from an array of values [0.0004, 0.004,0.008,0.02]) in the fitting. For the dust attenuation, we adopt the module, {\tt dustatt\_modified\_starburst} based on \citet{2000ApJ...533..682C} starburst attenuation curve. The input parameters for color excess or reddening of stellar continuum and nebular lines are provided in accordance with our dust attenuation calculation in previous section~\ref{sec:DE}. We fix the power law slope ($\delta$) of the dust attenuation curve to $-0.5$ which is steeper than the \citet{2000ApJ...533..682C} curve ($\delta$ = 0) and the UV bump amplitude to $1.0$, respectively. In addition, we use \citet{Dale2014} module to model polycyclic aromatic hydrocarbons emission. Under this module, we consider no AGN contribution and IR power law slope is set to $2.0$. The above mentioned modules and input parameters remain unchanged throughout the process.
\par
\begin{figure}
    \centering
    \includegraphics[width=\columnwidth]{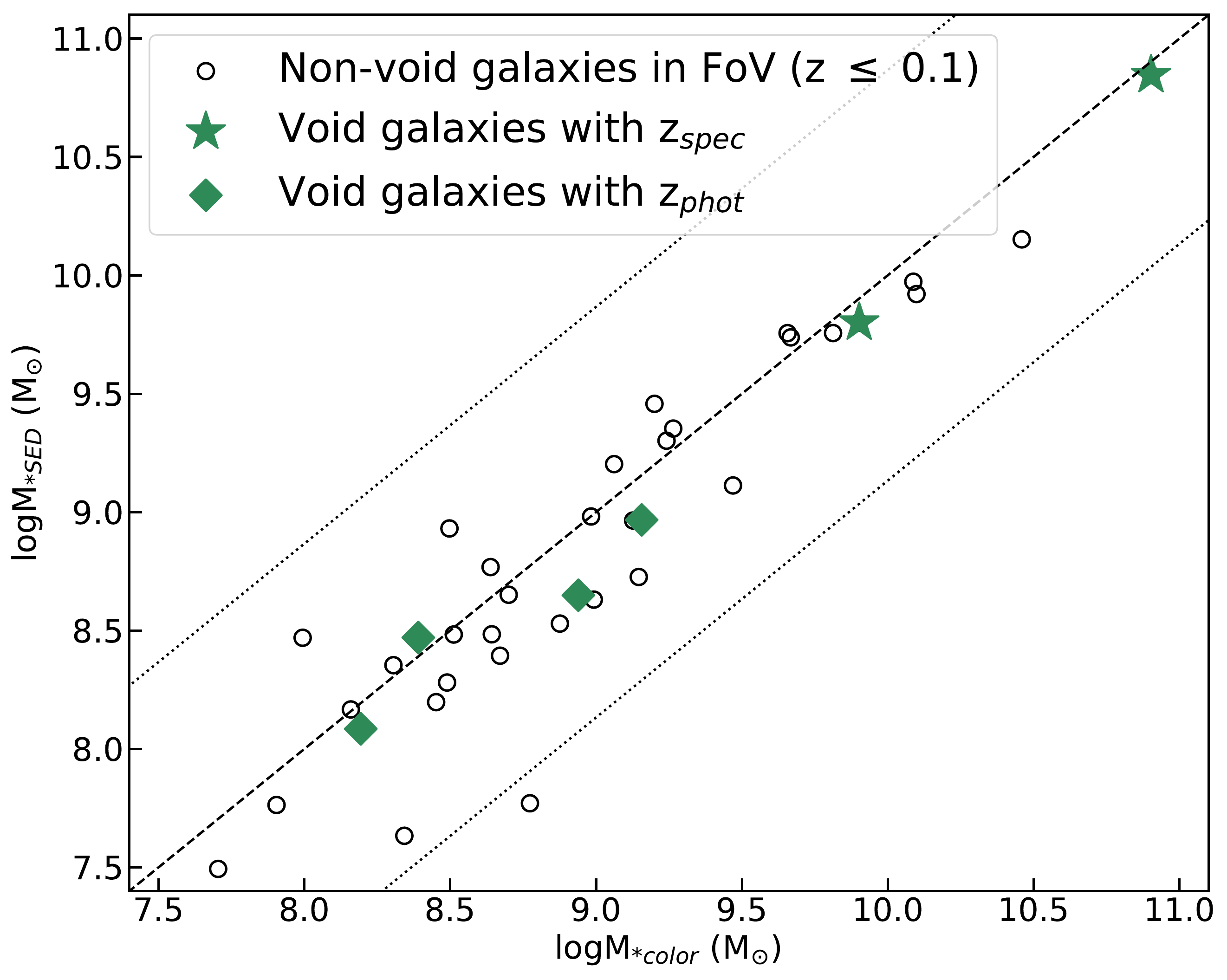}
    \caption{Comparison between SED-derived and color-derived stellar masses for void and non-void galaxies ($z$ $\leq$ 0.1) in our FoV. Dashed line represents one to one relation while dotted lines represents 1$\sigma$ scatter.}
    \label{fig:M*}
\end{figure}

In the second method, color-based stellar masses (M$_{*color}$) for individual galaxies are obtained from the relation between $g-r$ color and stellar mass-to-light ratio corresponding to optical luminosity (L$_\mathrm{r}$) using \citet{Bell2003} which is based on `diet' Salpeter IMF. Later we multiply M$_{*color}$ by a factor of $0.7$ to scale it to normal Salpeter IMF for appropriate comparison with SED-based stellar masses (M$_{*SED}$). Figure~\ref{fig:M*} shows one to one relation between M$_{*SED}$ and M$_{*color}$. Both set of stellar masses are in close agreement with each other. Henceforth, we use M$_{*SED}$ throughout the work for analysis. Void galaxies with z$_{phot}$ reported in the work are low-mass systems (M$_*$ $\lesssim$ 10$^{9}$ M$_{\odot}$) and evidently, M$_*$ for most the void galaxies lies below the stellar mass of a L$_{*}$ galaxy, i$.$e$.$, $3 \times 10^{10}$ M$_{\odot}$.

\medskip
\section{FUV STAR FORMATION RATE}
\label{sec:sfr}

The star formation rate provides key insight into the assembly history of a galaxy's stellar mass. Far-UV fluxes emitted by young, massive stars (typically O, B type) amounts to the instantaneous star formation in a galaxy. In other words, far-UV fluxes (if internal extinction corrected) can provide one of the best estimates of the recent star formation (over $\sim 100$ Myr) in a galaxy. The FUV emission and the associated SFR has been estimated in galaxies with different Hubble types ranging from late-types to early-types \citep{calzetti2012star,2005ApJ...619L.111Y}. We have calculated the FUV star formation rate (in units of M$_{\odot}$yr$^{-1}$) using the following relation given by \citet{1998ARA&A..36..189K}. 

\begin{equation}
SFR_\mathrm{FUV}=1.4 \times 10^{-28} L_\mathrm{FUV} (ergs\ s^{-1}\ Hz^{-1})
\end{equation}

\noindent Where L$_{\mathrm{FUV}}$ is the intrinsic FUV luminosity of a galaxy. The FUV SFRs are calibrated assuming that the star formation history of a galaxy is constant for the last $\sim 100$ Myr. In Table~\ref{tab:void_galaxies}, we show the SFR along with UV magnitudes (FUV/NUV), stellar masses, absolute magnitudes (M$_\mathrm{r}$), optical color g$-$r, and UV$-$optical color NUV$-$r for our sample of void galaxies. The FUV SFRs for the void galaxies detected in the FoV spans a wide range from 0.05 M\textsubscript{\(\odot\)}yr$^{-1}$ to 51.01 M\textsubscript{\(\odot\)}yr$^{-1}$ with median SFR$_\mathrm{FUV}$ $\sim$ 3.96 M\textsubscript{\(\odot\)}yr$^{-1}$. The FUV SFRs for most of our sample galaxies are comparable to that of a normal spiral galaxy within the local volume and are higher than that of a low-mass, star-forming dwarf galaxy.  
\par
\begin{figure}
    \centering
    \includegraphics[width=\columnwidth]{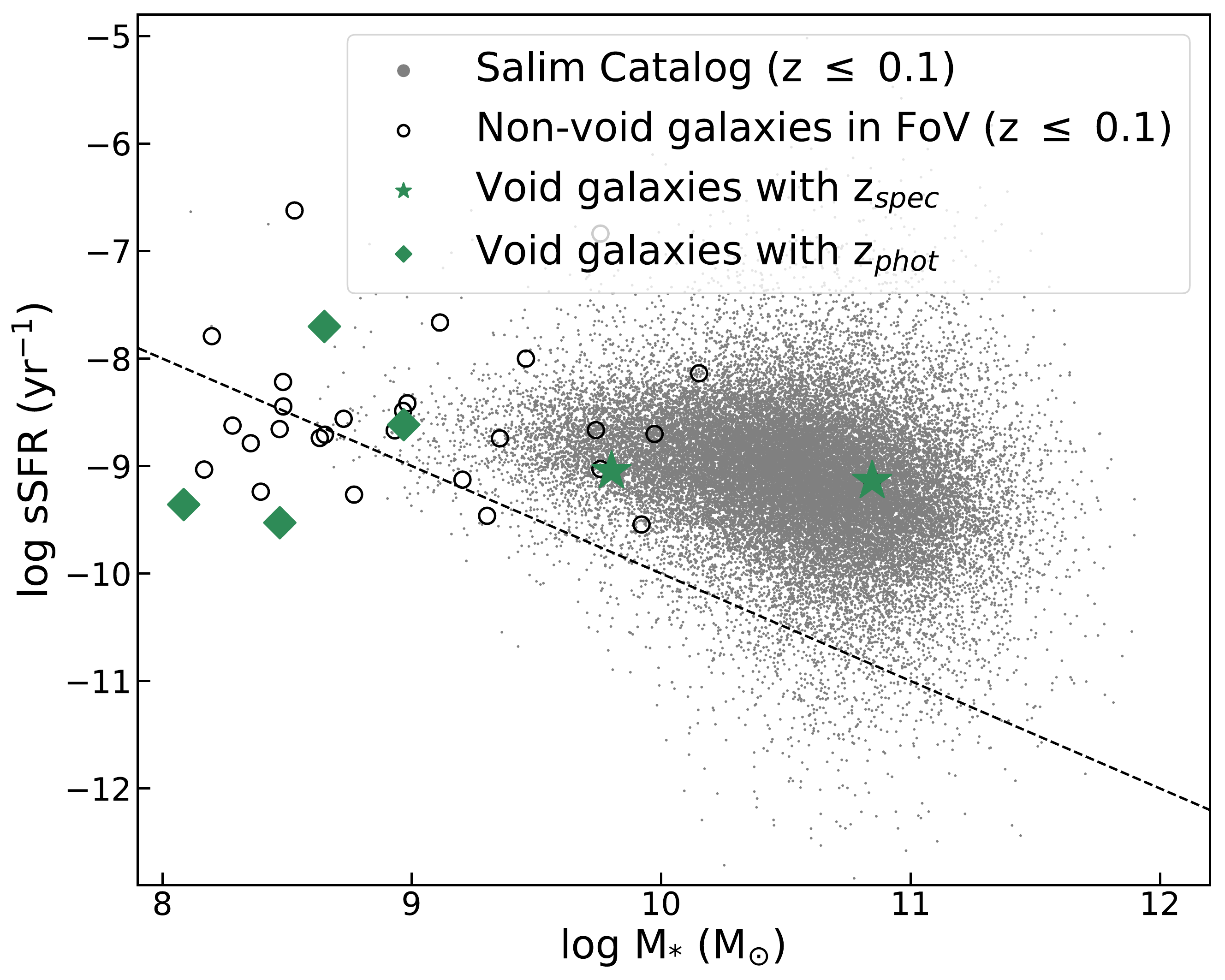}
    \caption{FUV sSFR vs$.$ Stellar mass for void and non-void galaxies ($z$ $\leq$ 0.1). The trend of local galaxies shown using $Salim\text{ }Catalog$ in the background. Dotted line represents galaxies with SFR = 1 M$_{\odot}$yr$^{-1}$.}
    \label{fig:ssfr}
\end{figure}

In Figure~\ref{fig:ssfr}, we show the distribution of void and non-void galaxies on the FUV sSFR-M$_*$ plane. The background galaxies comprise of $Salim\text{ }Catalog$ with $z$ $\leq$ 0.1. We compute the internal dust corrected FUV SFRs and color-based stellar masses for the background sample using the same recipe as described in the preceding sections. The background galaxies from $Salim\text{ }Catalog$ are well distributed over the sSFR$-$M$_{*}$ plane. However, as Figure~\ref{fig:ssfr} shows, most of the void galaxies with photo-z lie on the low-mass end of the distribution and they are basically vigorously star-forming galaxies with $\log(\mathrm{sSFR})$ ranging from $-$9.5 yr$^{-1}$ to $-$7.7 yr$^{-1}$ with a median $\approx$ $-$9.09 yr$^{-1}$. These values signify that all the void galaxies detected in our work are star-forming in nature. Even the most massive galaxy in our sample belongs to the star-forming cloud. Interestingly, the sSFR for the non-void galaxies detected in our FoV are comparable to those of void galaxies.

\begin{deluxetable*}{ccccccccccc}
% \tablenum{2}
\tablecaption{Photometric details, SFR$_\mathrm{FUV}$ and stellar masses of six void galaxies reported in the work.\label{tab:void_galaxies}}
% \tablewidth{0pt}
\tablehead{
\colhead{S No.} & \colhead{RA} & \colhead{DEC} & \colhead{FUV$_\mathrm{AB}$} &
\colhead{NUV$_\mathrm{AB}$} & \colhead{$z$} & \colhead{SFR$_\mathrm{FUV}$} & \colhead{M$_{*}$} & \colhead{M$_\mathrm{r}$} & \colhead{g$-$r} & \colhead{NUV$-$r}\\
\colhead{} & \colhead{J2000} & \colhead{J2000} & \colhead{mag} &
\colhead{mag} & \colhead{} & \colhead{M\textsubscript{\(\odot\)} yr$^{-1}$} & \colhead{10$^{10}$ M\textsubscript{\(\odot\)}} & \colhead{mag} & \colhead{mag} & \colhead{mag} 
}
% %\decimalcolnumbers
\startdata
 G1 & 14:07:25.63 & 48:50:43.4 & 20.29$\pm$0.05 & 19.80$\pm$0.01 & 0.055$\pm$0.012 & 8.874 & 0.044 & $-$18.47 & 0.13 & 1.13 \\ 
 G3 & 14:08:43.44 & 48:54:10.8 & 22.27$\pm$0.14 & 22.47$\pm$0.05 & 0.043$\pm$0.011 & 0.053 & 0.011 & $-$16.00 & 0.35 & 1.84 \\ 
 G4 & 14:07:53.33 & 49:01:13.1 & 21.52$\pm$0.10 & 21.91$\pm$0.04 & 0.057$\pm$0.004 & 0.088 & 0.029 & $-$16.74 & 0.26 & 1.44 \\
 G6 & 14:07:40.08 & 49:05:10.5 & 20.88$\pm$0.07 & 20.58$\pm$0.02 & 0.055$\pm$0.014 & 2.253 & 0.093 & $-$18.44 & 0.34 & 1.91 \\ 
 S1 & 14:08:13.59 & 48:51:44.7 & 19.06$\pm$0.03 & 18.40$\pm$0.01 & 0.0518$\pm$0.0001 & 51.010 & 7.009 &$-$22.08 & 0.60 & 3.35 \\ 
 S2 & 14:08:11.40 & 48:53:44.4 & 19.36$\pm$0.04 & 19.15$\pm$0.01 & 0.0511$\pm$0.0002 & 5.668 & 0.631 & $-$20.16 & 0.39 & 2.30 \\
\enddata
\tablecomments{Colors and absolute magnitudes are K-corrected and extinction corrected. SFR$_{FUV}$ are also corrected for internal extinction. G1, G3, G4 and G6 - void galaxies with $z_\mathrm{phot}$; S1 and S2 - void galaxies with $z_\mathrm{spec}$.}
\end{deluxetable*}

%%%%%%%%%%%%%%%%%%%%%%%%%%%%%%%%%%%%%%%%%%%%%%

%%%%%%%%%%%%%%%%%%%%%%%%%%%%%%%%%%%%%%%%%%%
\section{COLOR-MAGNITUDE DIAGRAMS}
\label{sec:cmd}
In this section, we summarize the results from the UV/optical/NIR color-magnitude diagrams (CMDs) to study the properties of our sample void galaxies. Our void galaxies are divided in two categories, i$.$e$.$ with $z_\mathrm{spec}$ and z$_\mathrm{photo}$, based on the means of their redshift determination. The galaxies detected outside the Bootes Void having either redshifts (photometric/ spectroscopic) are termed as non-void galaxies with $z$ $\leq$ 0.1 in the subsequent CMDs. 

\subsection{UV color$-$magnitude diagram}
\label{subsec:UVc}
\begin{figure}
	\includegraphics[width=\columnwidth]{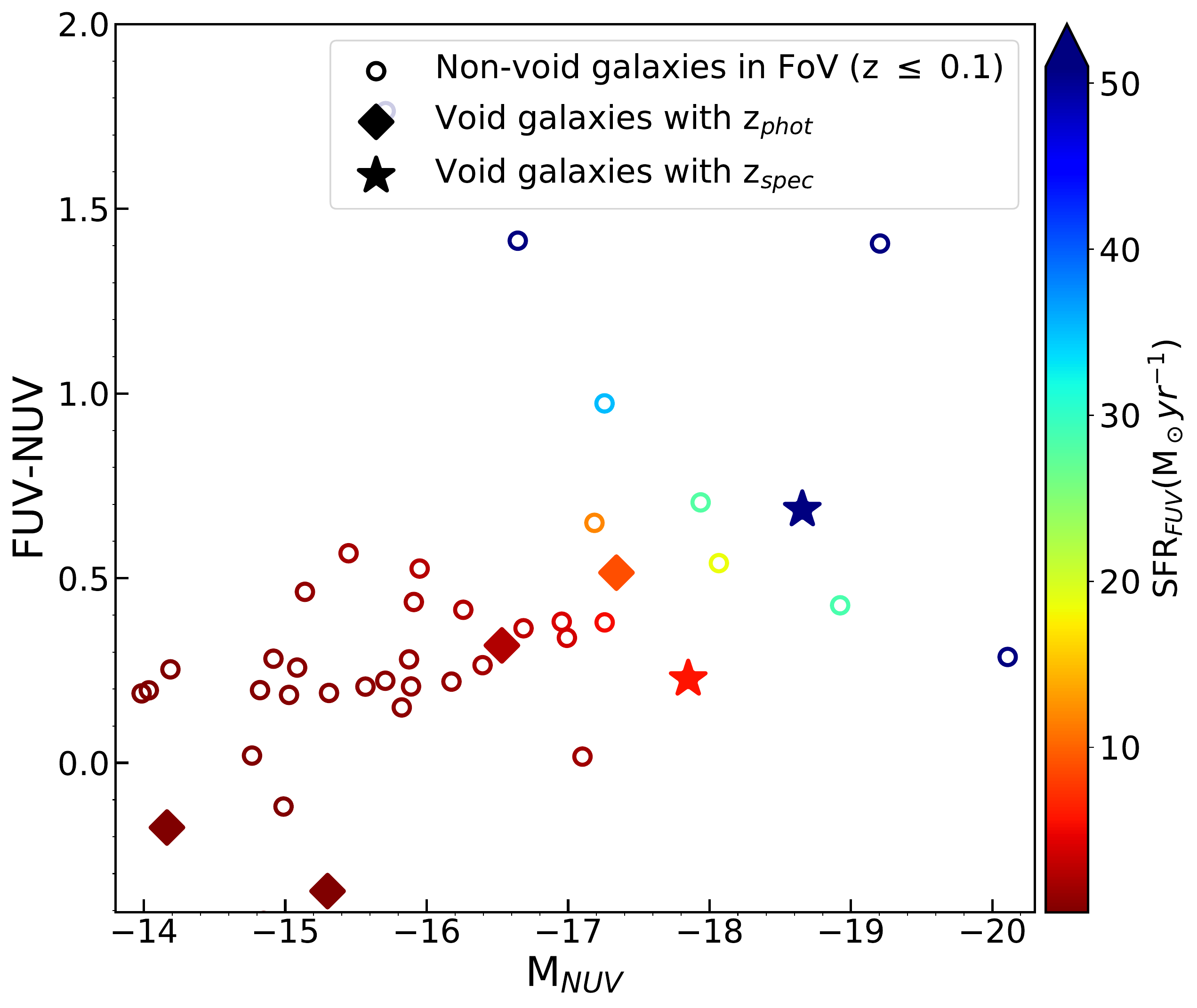}
    \caption{FUV$-$NUV vs$.$ M$_{\mathrm{NUV}}$ CMD for all galaxies (z $\leq$ 0.1) detected in the UVIT FoV. Each symbol represents a galaxy colour coded with FUV SFR. Filled diamonds are the void galaxies with z$_\mathrm{phot}$ detected in our FoV whereas filled stars denotes void galaxies with $z_{\mathrm{spec}}$. The open circles represent non-void galaxies detected in our survey.}
    \label{fig:UV}
\end{figure}

The FUV$-$NUV color for a large sample galaxies varies across $\sim$ 2 mag (see Figure~\ref{fig:UV}). In general, star-forming galaxies are found to have an average FUV$-$NUV color $\approx$ 0.4 mag and the color peaks at 0.9 mag where the transition from late (young) to early (old) type galaxies takes place \citep{2007ApJS..173..185G}. In Figure~\ref{fig:UV}, we have shown UV CMD distribution for all galaxies detected in our FoV upto $z$ ${\leq}$ 0.1 wherein the side color bar represents their internal dust corrected FUV SFRs. The FUV$-$NUV color of the void galaxies (with z$_\mathrm{spec}$ or z$_\mathrm{photo}$) are inclined towards the bluer end of the color scale with an average value of $\approx$ 0.2 mag indicating recent star formation in these systems along with late type or irregular morphological features. Based on the FUV$-$NUV colors and FUV SFRs, it is apparent that void galaxies comprise of a significant amount of young stellar population. However, we do not observe any strong correlation between FUV SFRs and FUV$-$NUV color for our entire sample of galaxies which is in agreement with \citet{Hunter_2010}. 
\par
\citet{2005ApJ...619L..15W} derived UV luminosity function for local galaxies ($z$ $\leq$ 0.1) for which the characteristic NUV magnitude (M$^{*}_{\mathrm{NUV}}$) came out to be $-$18.23 mag whereas M$_{\mathrm{NUV}}\text{ }\epsilon$  [$-$14.16, $-$18.65] mag for our sample of void galaxies implying that the distribution of our void sample traverses both the galaxy population type. Figure~\ref{fig:UV} shows no major difference in FUV SFRs of the void and non-void galaxies. Previously reported work such as \citet{2019ApJ...883...29W,2016MNRAS.458..394B,2008MNRAS.383.1058C} deduce similar results where impact of the environment on the SFRs of galaxies were found to be insignificant. However, the total fraction of blue/red galaxies is strongly dependent on the environment at a given stellar mass range \citep{blue_fraction,2006MNRAS.373..469B}.  
 
\subsection{UV$-$NIR color-magnitude diagram}

\begin{figure}
	\includegraphics[width=\columnwidth]{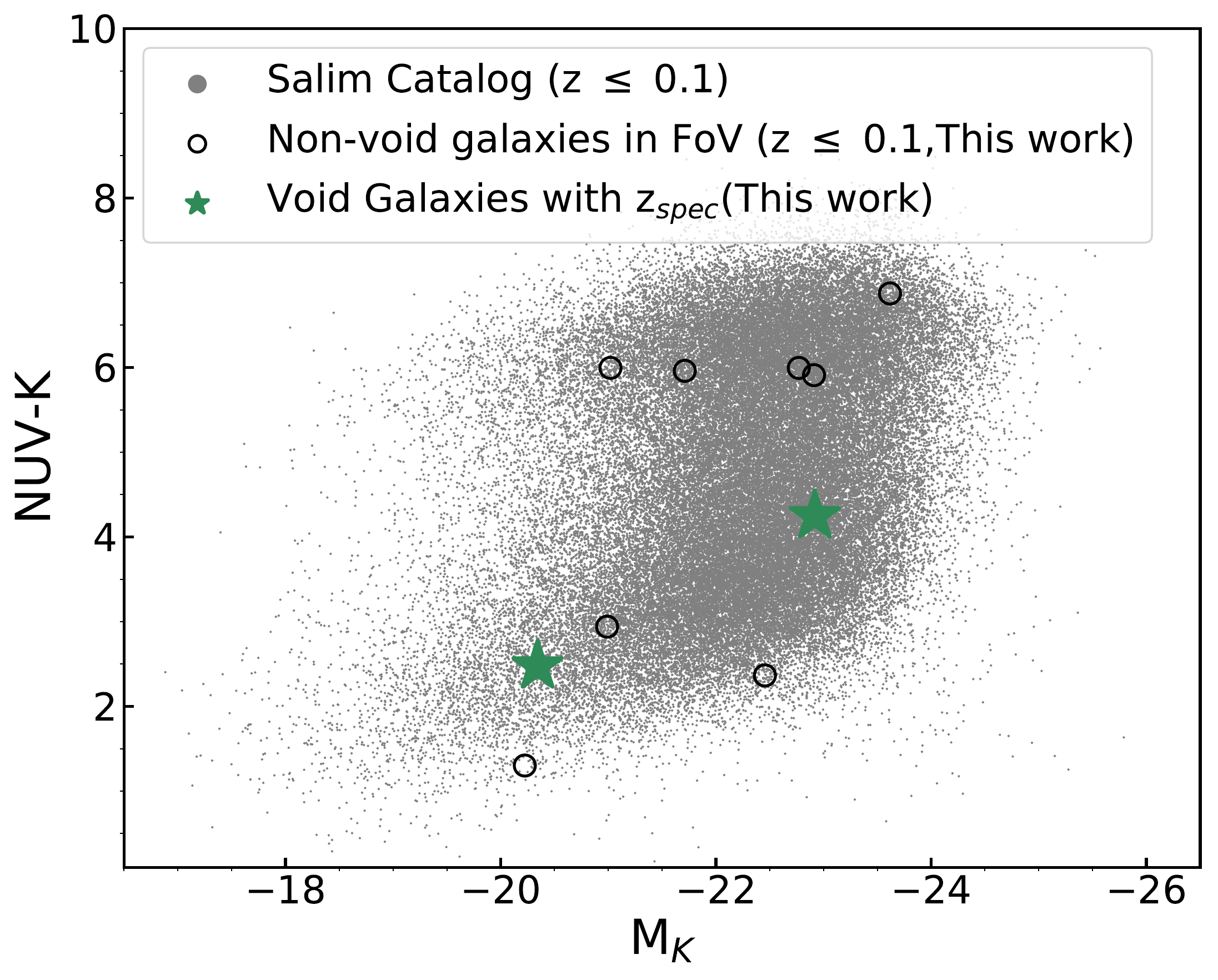}
    \caption{NUV$-$K vs$.$ M$_\mathrm{K}$ CMD of galaxies detected in UVIT. The background galaxies (grey circles) are from $Salim\text{ }Catalog$ upto ($z$ $\leq$ 0.1) \citep{2016ApJS..227....2S}. Black open circles represent galaxies residing outside the Bootes Void in our FoV ($z$ $\leq$ 0.1). Void galaxies with $z_{\mathrm{spec}}$ are shown by green star-shaped symbols }
    \label{fig:nuvk}
\end{figure}

In Figure~\ref{fig:nuvk}, the background galaxies are from $Salim\text{ }Catalog$ ($z$ $\leq$ 0.1). We refer to {\em 2MASS} all-sky Extended Source Catalog (XSC) \citep{Jarrett_2000} for procuring K-band magnitudes for all galaxies present in $Salim\text{ }Catalog$. The {\em 2MASS} XSC magnitudes are converted to AB magnitude system using the relation given in \citet{Blanton_2005}. On NUV$-$K vs$.$ M$_\mathrm{K}$ color$-$magnitude plane, the distribution of galaxies is bivariate as can be seen in Figure~\ref{fig:nuvk}. The NUV$-$K color provides a range of ${\approx}$8 mag which can be used efficiently to distinguish between galaxies based on their morphologies and stellar population type (early/late). Also, K$-$band luminosity is a tracer for total stellar mass of a galaxy \citep{Bell2003}.

\par As most of our photometrically verified void galaxies are absent in the NIR observations, therefore, we only study properties of void galaxies with spectroscopic observations using this CMD (see Figure \ref{fig:nuvk}). We scale (NUV$-$K)$_\mathrm{AB-Vega}$ color from \citet{2007ApJS..173..185G} to (NUV$-$K)$_\mathrm{AB-AB}$ magnitude system following prescriptions given by \citet{Blanton_2005} and find that the blue sequence comprising of spirals and irregular galaxies peak at 3.55 mag. The two void galaxies belong to the blue sequence as seen in Figure~\ref{fig:nuvk}. Here, the absolute magnitudes, M$_\mathrm{K}$ of these galaxies show a striking difference of two magnitude implying a significant variation in their total stellar masses. 

\subsection{Galaxy Bimodality using optical colors}

\begin{figure}
	\includegraphics[width=\columnwidth]{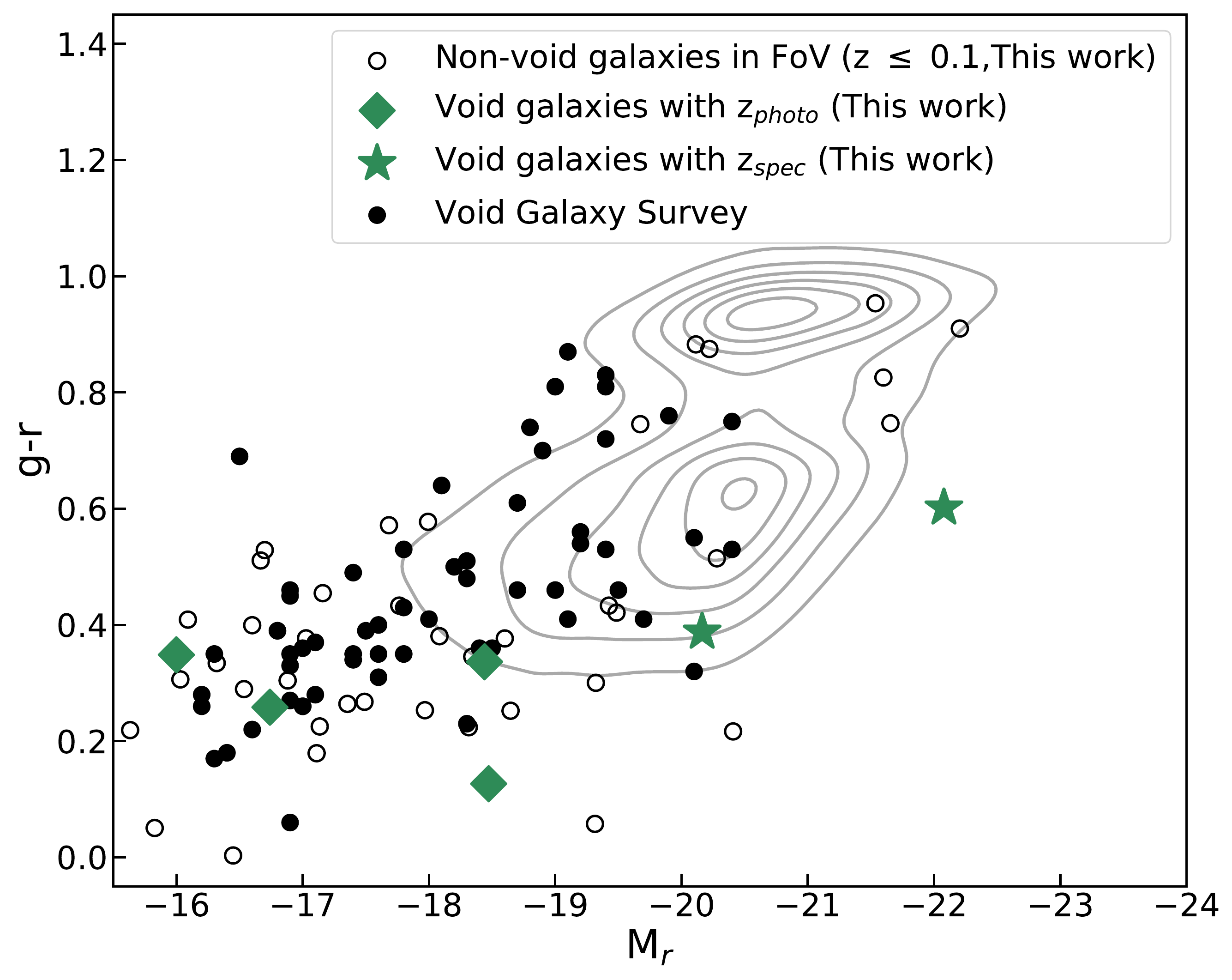}
    \caption{Comparison of g$-$r vs$.$ M$_\mathrm{r}$ CMD of our UVIT detected void galaxies with VGS (black filled circles) \citep{2012AJ....144...16K}. The constituent galaxies of the contour is taken from {\em SDSS} DR12 spectroscopic catalog with $z \leq  0.1$. Void galaxies with $z_{\mathrm{spec}}$ and $z_{\mathrm{photo}}$ are represented by green stars and diamonds, respectively. Black open circles represent the galaxies detected outside the void ($z$ $\leq$ 0.1). }
    \label{fig:SDSSbm}
\end{figure}

Optical colors have been quite successful in classifying galaxies in the local Universe \citep{Stratevaetal2001}. Galaxies present in local Universe can be broadly classified into two categories, i$.$e$.$, star formation quenched galaxies which are dominated with elliptical and S0s, likely to be found in denser environments and actively star-forming galaxies with spiral, disc-like and irregular morphologies mostly residing in the sparse environment \citep{Kauffmann_2004}. These galaxies tend to separate themselves into two groups based on UV$-$optical, optical$-$optical, UV$-$NIR colors up to $z$ $\sim$ 1 \citep{2004ApJ...600..681B,2005ApJ...619L.111Y,Wyder_2007}. In Figure~\ref{fig:SDSSbm}, we show g$-$r vs$.$ M$_\mathrm{r}$ color-magnitude distribution that is circumcentered around two modes: Blue Cloud peaking at g$-$r = 0.5 mag and Red Sequence peaks at g$-$r = 0.9 mag. Galaxies which fall in between the two groups are said to be Green Valley galaxies \citep{Salim2014}. 
\par
Figure~\ref{fig:SDSSbm} show optical CMD of UVIT identified void and non-void galaxies present in our FoV. In the background, we use a magnitude limited sample of 1,16,010 galaxies brighter than r $<$ 17.77 mag from {\em SDSS} upto $z$ ${\lesssim}$ 0.1 to construct the color magnitude contours. Nearly all our UVIT detected void galaxies belong to the Blue Cloud population, which fits the conventional understanding of galaxy formation and evolution. Thereby, the red counterpart of the bimodal distribution is unseen in our void sample. The two spectroscopically verified void galaxies belong to two different population type, i$.$e$.$, the Blue Cloud (image labelled as \textit{d} in Figure~\ref{fig:FOV}) and the Green Valley (image labelled as \textit{c} in Figure~\ref{fig:FOV}). The remaining void galaxies with z$_{\rm phot}$ are blue in color with late type morphologies. In totality, our sample follows a similar trend on the given optical CMD as shown by the galaxies present in Void Galaxy Survey (VGS) \citep{2012AJ....144...16K} (see Figure~\ref{fig:SDSSbm}). The absolute magnitudes, M$_\mathrm{r}$, for most of our sample and the VGS is fainter than ${\approx}$ $-$20 mag. A few of the galaxies from VGS are the members of the Red Sequence as seen in Figure~\ref{fig:SDSSbm}. However, we find none such galaxies for our sample. We observe that the non-void galaxies detected in our FoV belong to both the population type; spanning a wide range of optical color and luminosity while void galaxies majorly confine to the bluer and fainter end of the optical CMD.

\subsection{UV $-$ Optical color-magnitude diagram}

\begin{figure*}
    \centering
	\includegraphics[width=0.41\linewidth]{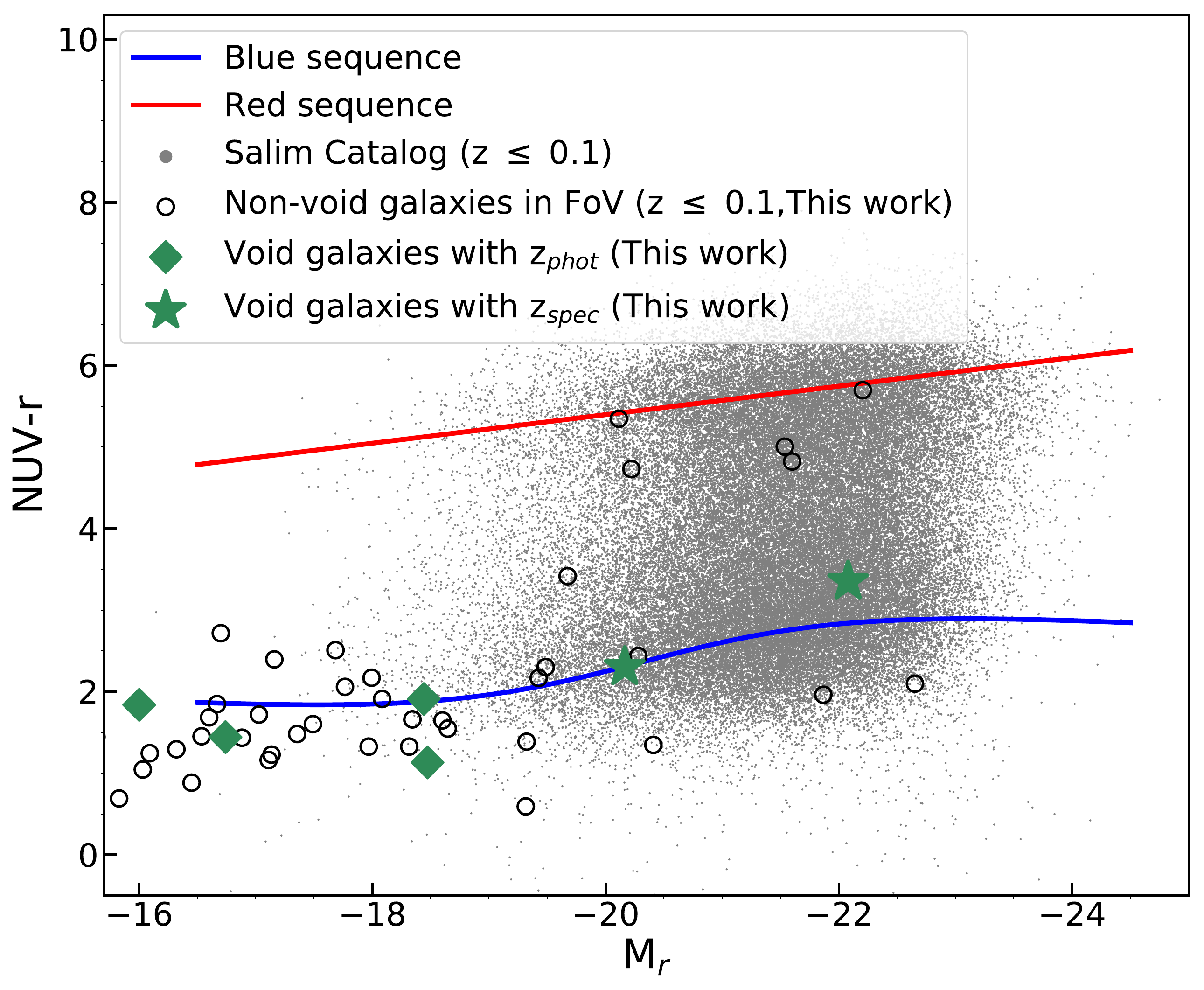}
	\includegraphics[width=0.45\linewidth]{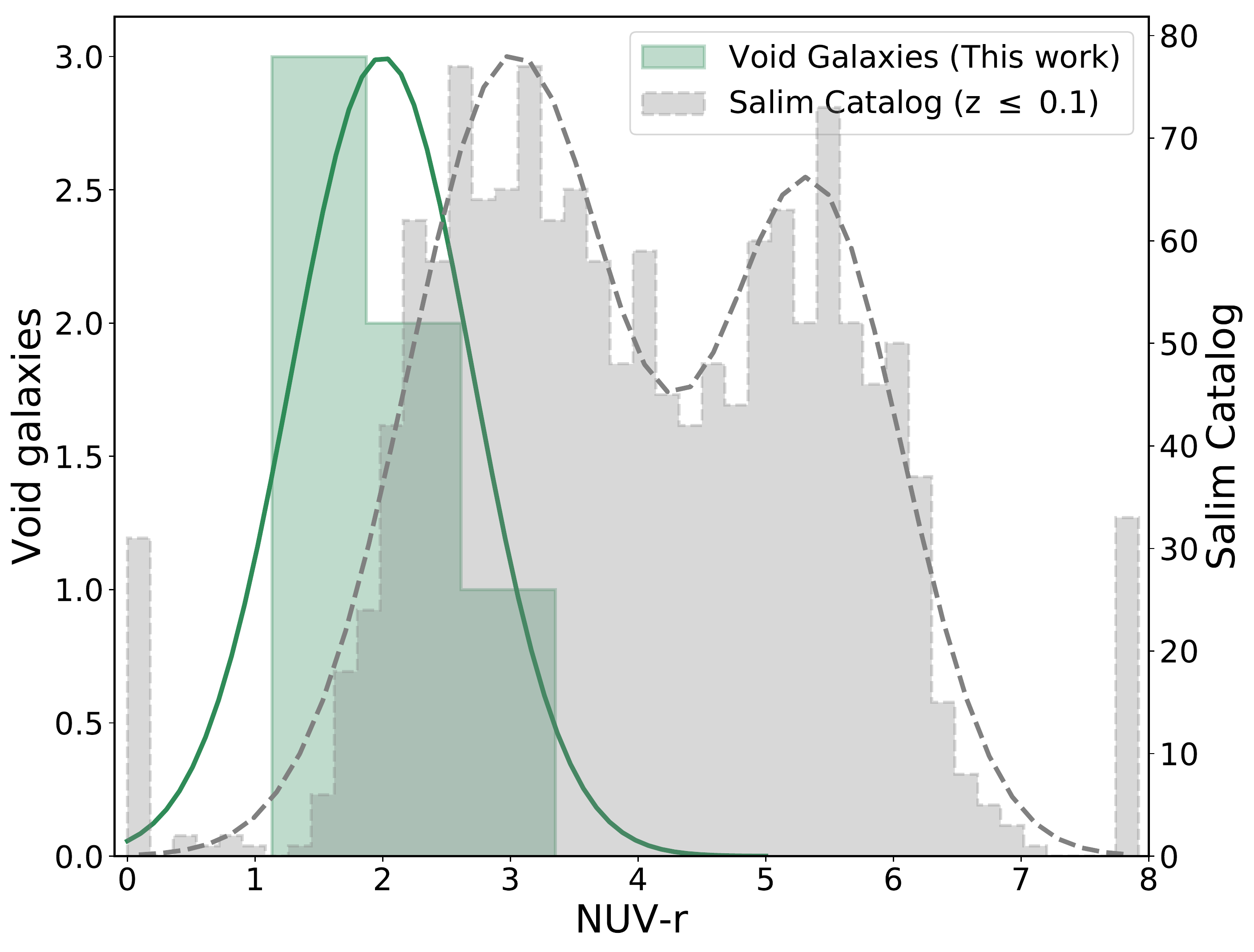}
    \caption{Right panel: NUV$-$r vs$.$ M$_\mathrm{r}$ CMD for galaxies in our FoV. Background galaxies with $z$ $\leq$ 0.1 taken from $Salim\text{ }Catalog$. Void galaxies with $z_{\mathrm{spec}}$ and $z_{\mathrm{photo}}$ are represented by green stars and diamonds, respectively. Non-void galaxies are marked by black open circles. Left Panel: NUV$-$r histogram for UVIT detected void galaxies and $Salim\text{ }Catalog$. The distributions are fitted with single peak and double peak Gaussian functions, respectively.}
    \label{fig:nuvr}
\end{figure*}

We have shown NUV$-$r vs$.$ M$_\mathrm{r}$ CMD for galaxies observed in our FoV in the left panel of Figure~\ref{fig:nuvr}. Background sample comprises galaxies in $Salim\text{ }Catalog$ ($z$ $\leq$ 0.1). The bivariate distribution of galaxies as a function of NUV$-$optical color and optical absolute magnitude is clearly visible. We fit the following relations to the peak color as a function of the absolute magnitude in the red sequence~\ref{eq:rpeak} and blue sequence~\ref{eq:bpeak}, respectively \citep{Wyder_2007}:

\begin{equation}
    (NUV - r) = 1.897- 0.175M_{r}
    \label{eq:rpeak}
\end{equation}

\begin{equation}
    (NUV - r) = 2.39 + 0.075(M_{r} + 20) - 0.808\tanh{\frac{M_r + 20.32}{1.81}}.
    \label{eq:bpeak}
\end{equation}

The UV$-$optical CMD have been extensively used in literature to follow the evolution of galaxies from the blue sequence to the red sequence, to study the evolution of early-type galaxies, and to deduce the mechanism responsible for star formation quenching \citep{Wyder_2007,Mazzei_2014,Kaviraj_2007}. The NUV$-$r color is a tracer of minimal amounts ($\sim$ 1\% mass fraction) of recent star formation ($\leq$ 1Gyr) (RSF) \citep{2007ApJS..173..512S}.   
\citet{Kaviraj_2007} suggest that galaxies with NUV$-$r $<$ 5.5 mag are likely to have undergone RSF confirming episodes of RSF for our void galaxies. The non-void galaxies in our FoV are distributed among both the population type, but we do not observe such a bimodality within the UVIT identified void galaxies.
\par
The NUV$-$r color histogram on the right panel of Figure~\ref{fig:nuvr} shows the color distribution for our sample and for galaxies present in $Salim\text{  }Catalog$. The galaxies from $Salim\text{ }Catalog$ show a clear bivariate distribution which fits well with a double peaked Gaussian function. The mean NUV$-$r colors for the blue and red sequences are $\mu_{Salim}^{blue} = 3.02$ mag and $\mu_{Salim}^{red} = 5.36$ mag, respectively whereas mean $\mu_{UVIT}^{VG}$ for our sample calculated by fitting a single component Gaussian profile equals $1.99$ mag. The spread in the NUV$-$r color for our UVIT detected void galaxies is unimodal, and centered below $\mu_{Salim}^{blue}$. Moreover, we perform Kolmogorov-Smirnov (KS) and Anderson-Darling (AD) tests on the NUV$-$r color distribution of our void galaxies and the blue sequence of $Salim\text{ }Catalog$ (NUV$-$r $\leq$ 4) to find whether the distribution of NUV$-$r color for the void galaxies are a subset of a larger sample of local galaxies. With p-value = $0.007$, high KS statistic (= $0.64$) and AD statistic (= $6.40$), we reject null hypothesis at a significance level = $0.05$ and infer that both sets of color belong to different parent populations. We acknowledge that our sample size for void galaxies is not significant enough for a strong statistical inference. The blue-ward shift in the NUV$-$r color of our void galaxies could be seen as a consequence of their low-density environment.
\par 
Intriguingly, we detect a few older (red) galaxies (NUV$-$r $>$ 4) outside the Bootes Void using UVIT observation, however, none of the void galaxies is seen to be passive, red and dead. Based on various CMDs studied in this work, we show that the star-forming void galaxies in our sample are fainter than their counterparts present in the field/ dense environment. Our sample of void galaxies lacks faint early-type galaxies such as dwarf ellipticals. Perhaps, one needs to have a dedicated, high-sensitivity infrared survey of galaxies in these sparse environments. 

%%%%%%%%%%%%%%%%%%%%%%%%%%%%%%%%%%%%%%%%%%%%%%%%%%%%%%%%%%%%
\section{Discussion and conclusions}
\label{sec:discuss}
The work primarily focuses on the photometric properties of the void galaxies detected in the Bootes Void, for which we have an ongoing survey covering a larger fraction of void using UVIT/{\em AstroSat}. The science-ready images are created first by processing the Level 1 data provided by ISRO using the official L2 pipeline. The end-product of this pipeline is an L2 image which is further corrected for astrometry. We use the appropriate {\em GALEX}-tiles and {\em SDSS} r-band images to correct for the astrometry in the L2 image (both in FUV and NUV). The difference in morphological features of a galaxy in various wavebands may have induced a slight offset ($\sim 0^{\prime\prime}.2 - 0^{\prime\prime}.3$) in the centroid (RA/ DEC) of sources in the final UVIT images (but see the color composite in Figure~\ref{fig:i+g+nuv}). 

\par
Most of the void galaxies reported by us lack spectroscopic observations. {\em SDSS} spectroscopic target selection criteria depend on the r-band apparent magnitude, and mean surface brightness \citep{sdss_spec} along with several other parameters. Our analysis and previous reports on void or isolated galaxies suggest that these systems have low optical luminosities and surface brightness \citep{2012AJ....144...16K,Hoyle_2005,Galaz_2011}. Therefore, one must reset the desired observational limits while surveying a void field. We encounter a few false detections and discrepancies in $STAR/ GALAXY$ classification in the archival {\em SDSS} photometric catalog. Hence, we perform $STAR/ GALAXY$ classification of our detected sources using UV-optical color-color diagrams. We work with {\em SDSS} photometric redshifts due to the absence of spectroscopic observation for all objects detected in our FoV. The error associated with {\em SDSS} photometric redshifts were significant enough to be included in our analysis. Thereafter, we use EAZY for determining photo-$z$ with better precision to assign void membership to the galaxies. Most of the galaxies with $z_\mathrm{phot}$ were either absent in {\em 2MASS} images or detected with poor SNR ($\lesssim$ 3). In the process, we only use photometric fluxes of seven wavebands. Hence, the lack of IR fluxes may induce slight inaccuracy in our photo-$z$ calculations. Spectroscopic observations of the final sample of four void galaxies with z$_\mathrm{phot}$ would confirm their candidature in the Bootes Void. In a similar manner, we exclude IR fluxes in the SED fitting process for determining M$_\mathrm{*SED}$ that may incur certain discrepancies in our calculations, although, we verify our results with M$_\mathrm{*color}$ and find good correspondence between the two stellar masses in most of the cases.
\par
UV emission from galaxies are subjected to extensive internal dust extinction. We calculate $A_\mathrm{FUV}$ with the help of two extreme UV broadband fluxes. This method tends to be erroneous as the UV continuum may get altered by some spectral features, and by the presence of old population \citep{Pilo}. Other techniques to calculate $A_\mathrm{FUV}$ require Balmer series line ratios - classic Balmer decrement method \citep{Grovesetal2012}, or total IR imaging observations \citep{Hao_2011} which are not available for our entire sample. 
\par
The work discusses about the effect of the global environment on the FUV SFRs and sSFRs of galaxies. The local affects such as galaxy interactions are not taken into account in our analysis. We argue that the global environment weakly impacts the ongoing star formation in galaxies which is supported by similar studies done previously. We stress on the fact that our sample size is small to provide a conclusive evidence to our findings. Quantities such as, SFRs and dispersion in sSFRs distribution depend on the stellar mass range of the galaxies taken under consideration \citep{Huang12,2012AJ....144...16K}. We further plan to investigate the problem with a large and diverse sample in terms of stellar mass for a concrete understanding of the environmental effects. Following are the primary scientific outcomes from our multi-wavelength analysis of star-forming galaxies present in Bootes Void:

\noindent{\bf 1.} We present a total of six void galaxies having FUV observation based on the deep UV imaging survey carried out by {\em AstroSat}/UVIT. Of these, three are new detections within the UVIT FoV. 
    
\noindent{\bf 2.} Our sample spans quite a range of stellar masses, even though, it is predominated by low-mass systems as most of them have stellar masses below L$_*$ galaxies.
    
\noindent{\bf 3.} The SFRs are corrected for the internal dust extinction using UV spectral slope $\beta$. The resultant values of $\beta$ suggest low to moderate dust obscuration in the void galaxies. 
    
\noindent{\bf 4.} The median SFR$_\mathrm{FUV}$ for the reported void galaxy sample $3.96$ M\textsubscript{\(\odot\)}yr$^{-1}$. The FUV SFRs of void galaxies are comparable to non-void low-mass, star-forming galaxies present in our sample. The ongoing moderate to high SFRs indicate the abundance of young massive O, B-type stars. Void galaxies show high values of sSFRs with median log(sSFR) $\approx$ $-9.09$ yr$^{-1}$ signifying on-going star formation at rapid timescales.
    
\noindent{\bf 5.} The UV, optical and NIR color-magnitude diagrams show that our void galaxies are bluer in color and possess disc-like, irregular morphologies, in some cases with spiral features. The most of our void galaxies have optical and UV luminosities less than L$^{\ast}$ galaxies. 

\noindent{\bf 6.} The color distribution of our void sample is confined to the blue sequence as seen in all the CMDs. In particular, we found a distinct shift in the NUV$-$r color distribution (Figure~\ref{fig:nuvr}) in our sample when compared to the blue sequence of a larger sample of local galaxies. This implies that galaxies present in voids are bluer than their counterpart present in the field or denser environment. 
    
\noindent{\bf 7.} Galaxies belonging to the red sequence are missing from our sample. Perhaps, a deeper infrared observation of the void region is in need to reach a firm conclusion. It could also be possible that a handful of galaxies in the low density environment are recently formed and are not matured yet. This remains to be investigated.   

\par    
\noindent {\bf Acknowledgements:}\\
We thank the referee for providing constructive suggestions/comments. The authors, DP and ACP thank Inter University center for Astronomy and Astrophysics (IUCAA), Pune, India for providing facilities to carry out this work. The UVIT project is a collaboration between IIA, IUCAA, TIFR, ISRO from Indian side and CSA from Canadian side. This publication uses UVIT data from the AstroSat mission of the Indian Space Research Organisation (ISRO), archived at the Indian Space Science Data Center (ISSDC). 

%The UVIT data used here was processed by the Payload Operations Centre at IIA. The UVIT is built in collaboration between IIA, IUCAA, TIFR, ISRO and CSA.

%This publication uses the data from the AstroSat mission of the Indian Space Research Organisation (ISRO). The UVIT data used here was processed by the Payload Operations Centre at IIA. The UVIT is built in collaboration between IIA, IUCAA, TIFR, ISRO and CSA. Complementary imaging of {\em UVIT} region is obtained from various independent observations including {\em GALEX}, {\em SDSS} DR12 and {\em 2MASS}.


%\vspace{5mm}
%\facilities{HST(STIS), Swift(XRT and UVOT), AAVSO, CTIO:1.3m,


\software{Astropy \citep{2013A&A...558A..33A},
          %DS9\citep{ds9},
          IRAF \citep{1993ASPC...52..173T},
          SExtractor \citep{1996A&AS..117..393B},
          EAZY \citep{brammer2008},
          CIGALE \citep{cigale}
          }

%\bibliographystyle{aasjournal}
%\bibliography{cite}

\end{document}